\documentclass[a4paper,11pt]{article}

\usepackage{jheppub} 
\usepackage{lineno}
\usepackage{txfonts}
\usepackage{physics}
\usepackage{subcaption}
\usepackage{mathrsfs}
\usepackage{amsmath}
\usepackage{tikz}
\usetikzlibrary{arrows,matrix,positioning}
\usepackage{appendix}

\title{Thermodynamics and kinetics of state switching for the asymptotically flat black hole in a cavity}
\author[a,*]{Ran Li, \note[*]{Corresponding author.}}
\author[b,*]{Jin Wang \note[*]{Corresponding author.}}
\affiliation[a]{Department of Physics, Qufu Normal University, Qufu, Shandong 273165, China}
\affiliation[b]{Department of Chemistry, and Department of Physics and Astronomy, The State University of New York at Stony Brook, Stony Brook, NY 11794, USA}
\emailAdd{liran@qfnu.edu.cn}
\emailAdd{jin.wang.1@stonybrook.edu}

\abstract{We propose that the thermodynamics and the kinetics of state switching for the asymptotically flat black hole enclosed by a cavity can be studied in terms of the free energy landscape formalism. The generalized free energy for the black hole enclosed by a cavity in the canonical ensemble is derived by using the York's approach, where the temperature on the cavity and the charges inside the cavity are kept as the fixed parameters. By quantifying the corresponding free energy landscape, we obtain the phase diagrams for the black hole in cavity, which reveal a Hawking-Page type transition for the uncharged black hole and a Van der Waals type transition for the charged black hole. We further assume that the dynamics of black hole state switching is mutually determined by the gradient force and the stochastic force arising from the free energy landscape and the thermal noises respectively. We then derive a recurrence relation for the $n$-momentum of the first passage time distribution function, which enables the calculation of the kinetic times characterized by the mean first passage time and its relative fluctuation. Our analysis illustrates that the kinetics of black hole state switching is determined by the ensemble temperature and the barrier height on the free energy landscape.}

\begin{document} 
\maketitle

\section{Introduction}

Since the discovery of Hawking radiation \cite{Hawking:1975vcx}, the thermodynamics of black holes has gained significant attention. It is generally believed that understanding the thermal features of black holes can promote the construction of a completed theory of quantum gravity.

Phase transition is a important topic in the field of the black hole thermodynamics. Great efforts have been made to understand the thermodynamics of the black hole phase transitions. In this aspect, two important examples are Hawking-Page phase transition \cite{Hawking:1982dh} and the Van der Waals type phase transition of the charged AdS black holes \cite{Chamblin:1999hg,Chamblin:1999tk}. Particularly in recent years, there has been considerable interest in the criticality and the phase transitions of asymptotically AdS black holes in the extended phase space \cite{Kubiznak:2012wp,Altamirano:2013uqa,Wei:2019uqg,Wei:2022dzw}, which is achieved by treating the cosmological constant as the thermodynamic pressure \cite{Kastor:2009wy,Dolan:2010ha,Dolan:2011xt}. It's noteworthy that these examples primarily focus on the asymptotically AdS black holes.

For the asymptotically flat black holes, York \cite{York:1986it} realized that Schwarzschild black holes placed inside a spherical cavity at finite radius can constitute a statistical ensemble with the cavity playing the role of the thermal reservoir. In this approach, the system has two stationary points that correspond two branches of Schwarzschild black holes \cite{Whiting:1988qr}. It is shown that the phase transition of Schwarzschild black holes in a cavity is similar to the Hawking-Page phase transition for the Schwarzschild AdS black holes \cite{York:1986it}. Using York’s approach, the Reissner-Nordstr\"{o}m black hole enclosed by a cavity in the grand canonical ensemble, where the temperature and the electric potential are specified at the boundary, was further discussed in \cite{Braden:1990hw}. The phase transition and critical behavior of the charged black holes in a cavity was latter studied in \cite{Carlip:2003ne,Lundgren:2006kt,Basu:2016srp}, where it is found that there is a first-order phase transition that terminates at a critical point. These behaviors are closely analogous to the phase behaviors for the charged AdS black holes \cite{Chamblin:1999hg,Chamblin:1999tk}. Note that the York's approach has also been generalized to study the thermodynamics and the phase transitions of asymptotically AdS and dS black holes \cite{Brown:1994gs,Peca:1998cs,Mitra:1999ge,Wang:2001gt,Wang:2002nq,Simovic:2018tdy,Wang:2019cax,Wang:2020hjw,Haroon:2020vpr,Banihashemi:2022jys,Jacobson:2022gmo,Jacobson:2022jir,Draper:2022ofa,Lemos:2024sjs}, as well as the black brane configurations in string theory \cite{Lu:2010xt,Lu:2013nt}, black holes in Gauss-Bonnet gravity \cite{Wang:2019urm,Peng:2020zmu,Su:2021jto,Marks:2021fpe}, hairy black holes in cavity \cite{Peng:2017gss}, Born-Infeld-de Sitter black holes \cite{Simovic:2019zgb,Liang:2019dni}, black holes in Teitelboim-Jackiw gravity \cite{Lemos:1996bq}, and black holes in higher dimensions \cite{Andre:2020czm,Andre:2021ctu,Fernandes:2023byx}.

Within the York's approach, the gravitational action is calculated by only imposing the Hamiltonian constraint condition on the spacetime geometry, which gives rise to the generalized free energy \cite{Whiting:1988qr}. In this regards, not only the on-shell black holes but also the off-shell black holes are taken into account to formulate the thermodynamics ensemble. The off-shell black holes are the spacetime configurations that can be reached via thermal fluctuations and the state switching process can occur via passing through such off-shell configurations. Therefore, the York's approach to the thermodynamics of black holes also provides a natural way to study the black hole state switching process.

In this paper, we will study the kinetics of state switching for the asymptotically flat black holes enclosed by a cavity in terms of the free energy landscape framework \cite{Li:2020khm,Li:2020nsy}. We give a rigorous derivation of the generalized free energy for the black hole with both electric charge and magnetic charge from the gravitational action by using the York's approach, where the temperature on the cavity and the charges in the cavity are kept as the fixed parameters. It is shown that the gravitational action is evaluated only by imposing the Gauss-law constraint and the gravitational Hamiltonian constraint. The full Einstein equations are not fully satisfied, which admits the existence of the off-shell spacetime configurations. By quantifying the corresponding free energy landscape, we study the phase transitions and phase structures for the black holes in cavity, which reveals a Hawking-Page type transition for the uncharged black hole and a Van der Waals type transition for the charged black hole.

We further assume that the dynamics of black hole state switching is determined by the Langevin equation where the gradient force and the stochastic force are originated from the free energy landscape and the thermal noises respectively \cite{Li:2021vdp}. For calculating the kinetic times that characterized the black hole state switching process, we derive a recurrence relation for the $n$-momentum of the first passage time distribution function. This enables the analytical expressions of the kinetic times characterized by the mean first passage time and its fluctuation. The numerical results for the kinetic times as the functions of ensemble temperature are calculated from the analytical expressions. Our analysis illustrates that the kinetics of black hole state switching is determined by the ensemble temperature and the barrier height on the free energy landscape.

This paper is arranged as follows. In Sec.\ref{rev_bh}, we briefly review the thermodynamics of the asymptotically flat black hole with both the electric charge and the magnetic charge. In Sec.\ref{cal_GFE}, we obtain the generalized free energy of the black hole-cavity system by evaluating the Euclidean gravitational action. In Sec.\ref{Thermo_phase_diagram}, we discuss the phase transitions and the phase diagrams for the uncharged black holes as well as the charged black holes. In Sec.\ref{kinetics_state_switching}, we first discuss how to describe the dynamical process of the black hole state switching, then derive the analytical expressions of the mean first passage time and its fluctuation, and finally present the numerical results for the kinetic times. The conclusion and discussion are presented in the last section.

\section{Asymptotically flat charged black hole}
\label{rev_bh}

In this section, we briefly discuss the geometry and the thermodynamics of the asymptotically flat black holes with the electric and the magnetic charges. This type of black hole is usually called the dyonic black hole, which was firstly proposed by Carter in \cite{Carter}\note[*]{This paper was reprinted in \cite{Carter:2009nex}.}. These black holes are solutions to the Einstein's field equations of general relativity coupled with electromagnetic field by considering the existence of magnetic monopoles. In the framework of string theory and supergravity, the high dimensional dyonic black holes are extensively investigated \cite{Rasheed:1995zv,Campbell:1992hc,Cheng:1993wp,Lu:2013ura,Dutta:2013dca,Jiang:2023gas}.

In the present work, we are interested in the four dimensional dyonic black hole, which the metric is given by 
\begin{eqnarray}\label{dyonic_metric}
    ds^2=-f(r)dt^2+\frac{1}{f(r)}dr^2+r^2\left(d\theta^2+\sin^2\theta d\phi^2\right)\;.
\end{eqnarray}
The blackening factor $f(r)$ is given by
\begin{eqnarray}\label{f_fun}
    f(r)=1-\frac{2M}{r}+\frac{Q^2+P^2}{r^2}\;.
\end{eqnarray}
Compared to the Reissner-Nordstr\"{o}m black hole, the spatial component of the gauge potential is opened to take the magnetic charge into account. The electromagnetic gauge potential is given by
\begin{eqnarray}
    A=-\frac{Q}{r}dt-P \cos\theta d\phi\;,
\end{eqnarray}
with $Q$ and $P$ being the electric and the magnetic charges respectively. 

The event horizon $r_+$ is determined by the largest root of the equation $f(r)=0$, which gives the analytical expression as
\begin{eqnarray}
 r_+=M+\sqrt{M^2-Q^2-P^2}\;.
\end{eqnarray}
In terms of the horizon radius $r_+$, the mass, the Hawking temperature and the entropy of the dyonic black hole are given by 
\begin{eqnarray}
    M&=&\frac{r_+}{2}\left(1+\frac{Q^2+P^2}{r_+^2}\right)\;,\nonumber\\
    T_H&=&\frac{r_+^2-(Q^2+P^2)}{4\pi r_+^3}\;,\nonumber\\
    S&=&\pi r_+^2\;.
\end{eqnarray}
It is also easy to check that these quantities satisfy the thermodynamic first law 
\begin{eqnarray}
    dM=T_HdS+\Phi_Q dQ+\Phi_P dP\;,
\end{eqnarray}
with $\Phi_Q=\frac{Q}{r_+}$ and $\Phi_P=\frac{P}{r_+}$ being the electric and magnetic potentials at the horizon. Furthermore, they satisfy the Smarr relation as 
\begin{eqnarray}\label{Smarr_eq}
    M=2T_HS+\Phi_Q Q+\Phi_P P\;.
\end{eqnarray}
These equations summarize the basic thermodynamic properties of the dyonic black holes. In the next section, we will employ the York's approach \cite{York:1986it,Whiting:1988qr,Braden:1990hw} to study the thermodynamics of the dyonic black hole enclosed by a cavity. 

\section{Generalized free energy from gravitational path integral}
\label{cal_GFE}

In this section, we derive the generalized free energy for the dyonic black holes enclosed in cavity by using the Euclidean gravitational path integral method \cite{Gibbons:1976ue}.

We start with the general form of the static spherically symmetric black hole metric in Euclidean signature \cite{Whiting:1988qr,Braden:1990hw} 
\begin{eqnarray}\label{metric_ansatz}
    ds^2=b^2d\tau^2+a^{-2}dr^2+r^2(d\theta^2+\sin^2\theta d\phi^2)\;,
\end{eqnarray}
where $\tau=it$ is the Euclidean time to preserve the positivity of the metric, and $b$ and $a$ are the functions of the radial coordinate $r$. The Euclidean time $\tau$ is assumed to have the period $2\pi$. The event horizon is given by $r=r_+$. The spacetime manifold $\mathcal{M}$ described by the metric \eqref{metric_ansatz} is bounded by a boundary $\partial\mathcal{M}$, which is represented by a cavity located at $r=r_B$. Therefore, the radial coordinate $r$ is bounded by $r_+\leq r \leq r_B$. In the present work, we will consider the canonical ensemble, where the temperature of the cavity is selected to be fixed. In addition, the electric and the magnetic charges enclosed in the cavity are also kept fixed. In this sense, the cavity plays the role of the thermal bath.

According to the argument made in \cite{York:1986it,Whiting:1988qr,Braden:1990hw}, the inverse temperature of the cavity $\beta$ is given by the proper length of the time coordinate at the boundary $r=r_B$
\begin{eqnarray}
    \beta=\int_0^{2\pi} b(r_B) d\tau=2\pi  b(r_B)\;,
\end{eqnarray}
which is just the inverse temperature of the event horizon redshifted to the cavity.

The ``center" or the inner boundary of the geometry is at the event horizon $r=r_+$ of the Euclidean black hole. Since there is a horizon, we must have 
\begin{eqnarray}
    b(r_+)=0\;.
\end{eqnarray}
The inner boundary of the Euclidean geometry is also required to be regular. Thus, each $\tau-r$ plane should be a two dimensional disk in the product manifold. By calculating the Euler characteristic number of the $\tau-r$ sector of the metric Eq.\eqref{metric_ansatz} in the Appendix \ref{Euler_2d}, we obtain the regularity condition as 
\begin{eqnarray}\label{bc1_inner}
     \left.(ab')\right|_{r=r_+}=1\;.
\end{eqnarray}
In addition, to distinguish the "hot flat space" with the topology $S^1\times R^3$ from the black hole topology $D^2\times S^2$, we calculate the Euler characteristic number of the four dimensional manifold with the boundary in Eq.\eqref{metric_ansatz} in the Appendix \ref{Euler_4d}, which gives us another boundary condition at $r=r_+$ as 
\begin{eqnarray}\label{bc2_inner}
     a(r_+)=0\;.
\end{eqnarray}
These boundary conditions are adequate to determine the generalized free energy of the dyonic black hole in cavity meanwhile admitting the fluctuations beyond the stationary black hole state.

With the boundary $\partial\mathcal{M}$ fixed, the appropriate Euclidean action for the Einstein-Maxwell theory in the canonical ensemble is given by \cite{Braden:1990hw}
\begin{eqnarray}\label{E_H_action}
I_E=I_{bulk}+I_{surf}\;,
\end{eqnarray}
with
\begin{eqnarray}
    I_{bulk}&=&-\frac{1}{16\pi} \int_{\mathcal{M}}d^4x\sqrt{g} \left( R-F_{\mu\nu}F^{\mu\nu}\right)\;,\\
    I_{surf}&=&-\frac{1}{8\pi} \int_{\partial\mathcal{M}} d^3 x\sqrt{h} \left(K-K_0+2n_{\mu} F^{\mu\nu} A_{\nu}\right)\;,
\end{eqnarray}
Here, $h_{\mu\nu}=g_{\mu\nu}-n_{\mu}n_{\nu}$ is the induced metric on the boundary $\partial\mathcal{M}$ where $n$ is an outward pointing unit normal vector to $\partial\mathcal{M}$. In the Gibbons-Hawking boundary term $I_{surf}$, the trace of the extrinsic curvature is defined by $K=\nabla^{\mu}n_{\mu}$. The Gibbons-Hawking surface term is to preserve a well posed variation problem while the surface term for the electromagnetic field is to preserve the fixed charge boundary condition at the boundary.

We first evaluate the gravitational part of the action. The results for the scalar curvature and the extrinsic curvature are given by
\begin{eqnarray}
    R&=&-2\left[\frac{\left(r a^2\right)'-1}{r^2}+\frac{a}{r^2 b}\left(r^2 a b'\right)'\right]\;,\nonumber\\
    K&=&a\left(\frac{2}{r}+\frac{b'}{b}\right)\nonumber\\
    K_0&=&\frac{2}{r}\;,    
\end{eqnarray}
where the prime represents the derivative with respect to the radial coordinate $r$. Performing the integration, we can get the gravitational part of the action as
\begin{eqnarray}\label{g_part}
   && -\frac{1}{16\pi} \int_{\mathcal{M}}d^4x\sqrt{g}  R -\frac{1}{8\pi} \int_{\partial\mathcal{M}} d^3 x\sqrt{h} \left(K-K_0\right)\nonumber\\
    &=&\pi\int_{r_+}^{r_B} dr \frac{b}{a}\left(\left(ra^2\right)'-1 \right)+
    2\pi r_Bb(r_B) -2\pi r_B a(r_B) b(r_B)-\pi r_+^2\;.
\end{eqnarray}

For the electromagnetic part, we consider the Maxwell equation 
\begin{eqnarray}
    \nabla_{\mu} F^{\mu\nu}=0\;,
\end{eqnarray}
with the gauge potential ansatz as 
\begin{eqnarray}
    A=A_{\tau}(r) d\tau+ A_{\phi}(\theta) d\phi\;.
\end{eqnarray}
Here, $A_{\tau}(r)$ represents the electric part generated by the electric charge while $A_{\phi}(\theta)$ represents the magnetic part generated by the magnetic charge. The Maxwell equation results in two Gauss constraints as 
\begin{eqnarray}\label{Gauss_con_eq}
    \left(\frac{r^2 a }{b} A_{\tau}' \right)'=0\;,\;\;\;
    \partial_\theta\left(\frac{1}{\sin\theta} \partial_\theta A_\phi \right)=0\;.
\end{eqnarray}
The first equation can be partly integrated as 
\begin{eqnarray}
    \frac{r^2 a }{b} A_{\tau}'=-iQ\;,
\end{eqnarray}
where $Q$ is the electric charge enclosed in the cavity. 
The second equation gives the solution to the magnetic part of the gauge potential as 
\begin{eqnarray}\label{A_phi_fun}
    A_\phi=P(1-\cos\theta)\;,
\end{eqnarray}
where $P$ is the magnetic charge in the cavity and the constant term is introduced to preserve the regularity when $\theta=0$. Evaluating the electromagnetic part, we get the action as 
\begin{eqnarray}\label{e_part}
    &&\frac{1}{16\pi} \int_{\mathcal{M}}d^4x\sqrt{g} F_{\mu\nu}F^{\mu\nu}-\frac{1}{4\pi} \int_{\partial\mathcal{M}} d^3 x\sqrt{h} n_{\mu} F^{\mu\nu} A_{\nu}\nonumber\\
    &=&\pi \int_{r_+}^{r_B}dr \frac{b}{a}\left(\frac{P^2}{r^2}-\frac{Q^2}{r^2}\right)+2\pi i Q A_\tau(r_B)\nonumber\\
    &=&\pi \int_{r_+}^{r_B}dr \frac{b}{a}\left(\frac{P^2}{r^2}+\frac{Q^2}{r^2}\right)  \;,
\end{eqnarray}
where we have used the fact that 
\begin{eqnarray}
    A_{\tau}(r_B)=-i\int_{r_+}^{r_B} dr \frac{b}{a} \frac{Q}{r^2}\;.
\end{eqnarray}
Here, $A_{\tau}(r_+)$ is selected to be zero in order to preserve the regularity of the norm of the gauge potential at the horizon.

Next, we consider the gravitational Hamiltonian constraint, which is the $\tau-\tau$ component of the Einstein field equation 
\begin{eqnarray}\label{Ham_con_eq}
    G^{\tau}_{\tau}-8\pi T^{\tau}_{\tau}=0\;,
\end{eqnarray}
which gives us the equation 
\begin{eqnarray}\label{a_eq}
    (ra^2)'-1+\frac{Q^2+P^2}{r^2}=0\;.
\end{eqnarray}
The solution to this equation is given by
\begin{eqnarray}
 a^2(r)=1-\frac{C}{r}+\frac{Q^2+P^2}{r^2}\;,
\end{eqnarray}
where $C$ is the integral constant. Imposing the boundary at the horizon given in Eq.\eqref{bc2_inner}, we have
\begin{eqnarray}\label{a_fun}
    a^2(r)=\left(1-\frac{r_+}{r}\right)\left(1-\frac{Q^2+P^2}{r_+ r}\right)\;.
\end{eqnarray}

By substituting Eq.\eqref{a_eq} into Eq.\eqref{g_part}, one can see that the first term in the gravitational action just cancels the electromagnetic action properly, which gives the final result as 
\begin{eqnarray}\label{York_action}
    I_E&=& 2\pi r_Bb(r_B) -2\pi r_B a(r_B) b(r_B)-\pi r_+^2\nonumber\\
    &=& \beta r_B\left(1-\sqrt{\left(1-\frac{r_+}{r_B}\right)\left(1-\frac{Q^2+P^2}{r_+ r_B}\right)}\right)-\pi r_+^2\;.
\end{eqnarray}
This in turn gives us the generalized free energy of the black hole-cavity system as 
\begin{eqnarray}\label{Generalized_Free_energy}
    F=\frac{I_E}{\beta}=r_B\left(1-\sqrt{\left(1-\frac{r_+}{r_B}\right)\left(1-\frac{Q^2+P^2}{r_+ r_B}\right)}\right)-\pi T r_+^2\;,
\end{eqnarray}
where $T=\frac{1}{\beta}$ is the ensemble temperature on the cavity. This expression for the generalized free energy resembles the form of $E-TS$, where the quasilocal energy play the role of the energy. When $P=0$, this result recovers that for the charged Reissner-Nordstr\"{o}m black hole in a cavity in canonical ensemble \cite{Lundgren:2006kt}.

In deriving this result, we have used the Gauss-law constraints \eqref{Gauss_con_eq} for the electromagnetic field and the Hamiltonian constraint \eqref{Ham_con_eq} for the gravitational field. However, the full Einstein equations are not required to satisfy. The metric function $a(r)$ and the $\phi$ component of the Gauge potential $A_\phi$ are specified by Eq.\eqref{a_fun} and Eq.\eqref{A_phi_fun} respectively, while the metric function $b(r)$ and the $\tau$ component of the Gauge potential $A_{\tau}$ are not known. Therefore, the metric function $b(r)$ is an arbitrary function of the radial coordinate $r$, which satisfies the fixed boundary conditions at $r=r_+$ and $r=r_B$. Especially, near $r=r_+$, we have the asymptotic expansion for $b(r)$ as
\begin{eqnarray}
    b(r)=\frac{2r_+^2}{\sqrt{r_+^2-(Q^2+P^2)}} \left(1-\frac{r_+}{r}\right)^{1/2}+\lambda \left(1-\frac{r_+}{r}\right)^{3/2}+\cdots\;,
\end{eqnarray}
which is not consistent with the dyonic black hole metric given by Eq.\eqref{dyonic_metric} and \eqref{f_fun} due to the existence of the high order terms. This inconsistency can be regarded as a kind of quantum fluctuations near the stationary dyonic black hole solution. In this sense, the generalized free energy given by Eq.\eqref{Generalized_Free_energy} is considered to describe the fluctuating black hole as well. 

\section{Thermodynamics of phase transitions for the asymptotically flat black holes in cavity}\label{Thermo_phase_diagram}

\subsection{Charged dyonic black holes}

The solutions for a dyonic black hole in a cavity as a reservoir are given by finding the stationary points of the generalized free energy. It is convenient to introduce the non-dimensional variables by using the radius of the cavity as the characteristic length
\begin{eqnarray}
    \mathcal{F}=\frac{F}{r_B}\;,\;\;\;x=\frac{r_+}{r_B}\;,\;\;\;
    q=\frac{Q}{r_B}\;,\;\;\;p=\frac{P}{r_B}\;,\;\;\;\mathcal{T}=4\pi r_B T\;.
\end{eqnarray}
Then the rescaled generalized free energy is given by
\begin{eqnarray}
    \mathcal{F}=1-\sqrt{\left(1-x\right)\left(1-\frac{q^2+p^2}{x}\right)}-\frac{1}{4}\mathcal{T} x^2\;.
\end{eqnarray}
In this expression, the electric charge $q$, the magnetic charge $p$ and the cavity temperature $\mathcal{T}$ are fixed by the canonical ensemble. The only free parameter is the horizon radius $x$, which is bounded by $[q^2+p^2,1]$.

\begin{figure}
  \centering
  \includegraphics[width=8cm]{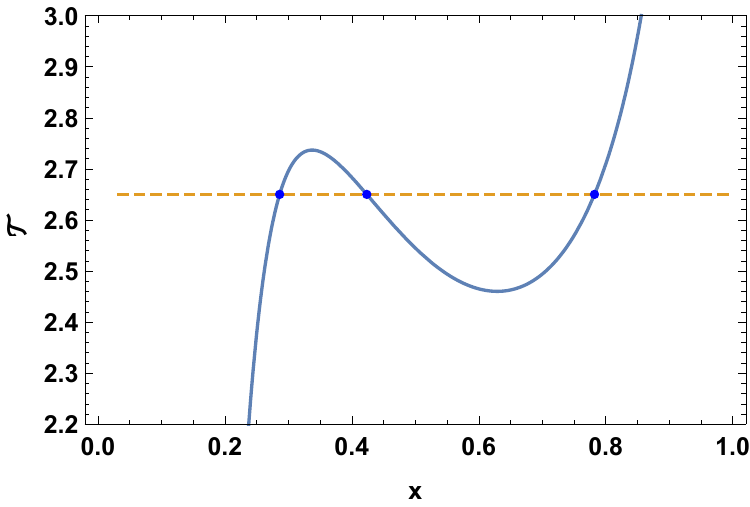}\\
  \caption{The temperature $\mathcal{T}$ as the function of black hole radius $x$. In this plot, $p=0.1$ and $q=0.15$. The horizontal dashed line represents $\mathcal{T}=2.65$. The blues dots represent the dyonic black holes in equilibrium states with the cavity. }
  \label{Temperautre_radius}
\end{figure}

Extremizing the generalized free energy with respect to $x$ gives us
\begin{eqnarray}\label{tem_eq}
    \mathcal{T}=\frac{\left(1-\frac{q^2+p^2}{x^2}\right)}{x\sqrt{\left(1-x\right)\left(1-\frac{q^2+p^2}{x}\right)}}\;.
\end{eqnarray}
This equation gives the equilibrium state condition for the dyonic black hole with the cavity \cite{Carlip:2003ne}. By solving this equation for the fixed parameters $q$, $p$ and $\mathcal{T}$ of the canonical ensemble, the equilibrium dyonic black hole can be determined by the event horizon $x$. We plot the temperature $\mathcal{T}$ as the function of black hole radius $x$ in Figure \ref{Temperautre_radius}. In general, it is a non-monotonic function. The intersections between the $\mathcal{T}(x)$ curve with the constant $\mathcal{T}$ gives the stationary points of the generalized free energy function. There is a temperature range $[\mathcal{T}_{min},\mathcal{T}_{max}]$ in which there are three stationary points for the system. For the case plotted in Figure \ref{Temperautre_radius}, we have $\mathcal{T}_{min}=2.46$ and $\mathcal{T}_{max}=2.74$. In the following, we will show that this temperature range will reduce to null when increasing the electric charge $q$.       

Now by keeping the magnetic charge $p$ as a fixed parameter, we want to show that there is a line of first order phase transitions in this region that terminates at a critical point in the $(\mathcal{T},q)$ plane. We will show that this critical point is the location of a second order phase transition.

\begin{figure}
  \centering
  \includegraphics[width=8cm]{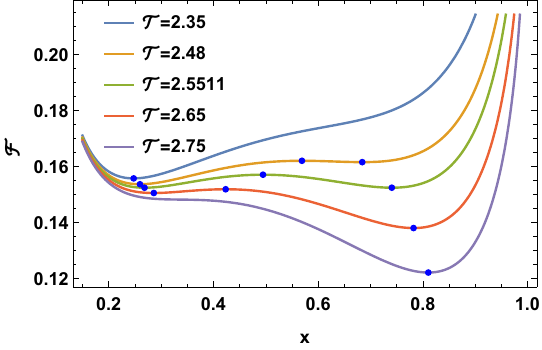}\\
  \caption{Free energy landscape for the charged black holes at different temperatures. In this plot, $p=0.1$ and $q=0.15$. The blue dots represent the dyonic black holes in equilibrium state with the cavity. }
  \label{Landscape}
\end{figure}

In Figure \ref{Landscape}, we have plotted the free energy landscapes at different temperatures. It shows that when $\mathcal{T}<\mathcal{T}_{min}$ and $\mathcal{T}>\mathcal{T}_{max}$, the landscape has the shape of single well, which means only one stationary point on it. When $\mathcal{T}_{min}<\mathcal{T}<\mathcal{T}_{max}$, the landscape has the double well's shape, which means there are three stationary points on it. In addition, when increasing the temperature, the small black hole state that is globally stable at the low temperature becomes the locally stable state at the high temperature. This implies that there exists a first order phase transition between the small black hole and the large black hole. In general, the temperature range $[\mathcal{T}_{min},\mathcal{T}_{max}]$ will be closed when increasing the charge $q$. This implies a second order phase transition at the end point.

\begin{figure}
  \centering
  \includegraphics[width=8cm]{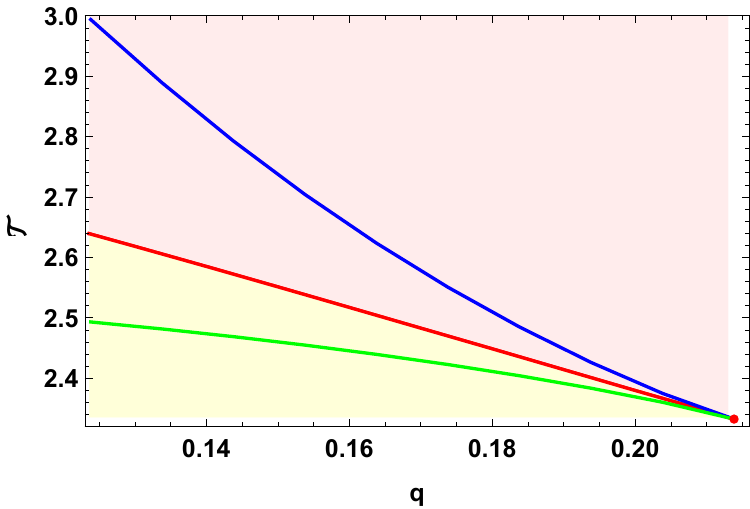}\\
  \caption{$\mathcal{T}-q$ phase diagram of the dyonic black hole in cavity for the fixed magnetic charge $p=0.1$. The first order phase transition is along the red line and terminates at the red point. }
  \label{Phase_Diagram}
\end{figure}

The simplest way to show this is to take the first order and the second order derivatives of generalized free energy $\mathcal{F}$ with respect to the black hole radius $x$. Recall that at the critical point, they are all equal to zero. Combine them to eliminate the temperature $\mathcal{T}$, and impose the condition that the resultant equation has two equal roots. It can be shown that the resultant equation is given by
\begin{eqnarray}\label{critical_condition}
   3 x^4 -2 \left(p^2+q^2+1\right) x^3 -6 \left(p^2+q^2\right) x^2 +6 \left(p^2+q^2\right) \left(p^2+q^2+1\right)  x-5 \left(p^2+q^2\right)^2=0\;.
\end{eqnarray}
The condition that this equation has two equal roots is that the discriminant of the above algebraic equation is zero, which is given by 
\begin{eqnarray}
    6912 \left(p^2+q^2-1\right)^4 \left(p^2+q^2\right)^3 \left(p^4+2 p^2 \left(q^2-9\right)+q^4-18 q^2+1\right)=0\;.
\end{eqnarray}
The solution to the above equation gives the critical value of the electric charge, which is given by
\begin{eqnarray}
    q_c=\sqrt{9-4 \sqrt{5}-p^2}\;.
\end{eqnarray}
It is clear that the critical value is dependent on the magnetic charge $p$. Substituting this critical value into Eq.\eqref{critical_condition}, one can get the critical value for the black hole radius as $x_c=5-2\sqrt{5}$. The corresponding critical temperature is $\mathcal{T}_c=\frac{2}{25} \left(5+2 \sqrt{5}\right)^{3/2}$. The phase diagram in $\mathcal{T}-q$ parameter space is presented in Figure \ref{Phase_Diagram}. The first order phase transition is determined by equal free energies of the two basins on the free energy landscape. By numerically solving the condition, one can get the red line on the phase diagram, which is also the coexistence curve of the small and the large black holes. In the light yellow region, the small black hole is thermodynamically stable while in the light pink region, the large black hole is stable. In the region between the blue and the green lines, the free energy landscape has the shape of double well, while outside of this region, there is only single well. 

At last, we point out that compared with the van der Waals type phase transition for the charged AdS black holes, the first order phase transition happens when the temperature is greater than the critical temperature $\mathcal{T}_c$.

\subsection{Uncharged black holes}

\begin{figure}
  \centering
  \includegraphics[width=8cm]{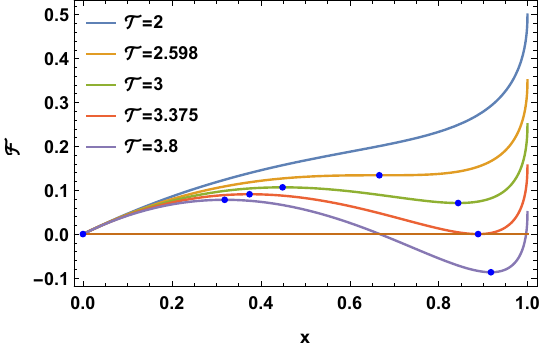}\\
  \caption{Free energy landscapes for the uncharged black hole at different temperatures. The blue dots represent the stationary black holes in equilibrium state with the cavity. }
  \label{Landscape_uncharged}
\end{figure}

For the uncharged case, we take $P=Q=0$. The corresponding generalized free energy is then given by 
\begin{eqnarray}
    \mathcal{F}=1-\sqrt{\left(1-x\right)}-\frac{1}{4}\mathcal{T} x^2\;.
\end{eqnarray}
The horizon radius $x$ is in the range $[0,1]$. The free energy landscapes for the uncharged black hole at different ensemble temperatures are plotted in Figure \ref{Landscape_uncharged}. Above the temperature $\mathcal{T}_{min}=2.598$, the free energy landscape has two branches of uncharged stationary solutions, one of which has the larger horizon radius and another has the smaller radius. Besides the black hole solutions, the system also admits the thermal vacuum solution without event horizon, i.e. $x=0$ solution denoted by the origin point. 
At the temperature $\mathcal{T}_c=3.375$, the thermal vacuum solution has the equal free energy with the large black hole solution, which means a phase transition between them. The behavior of the free energy landscapes with changing the ensemble temperature resembles that of the Hawking-Page phase transition discussed in \cite{Li:2020khm}. 

From Figure \ref{Landscape_uncharged}, one can see that When $\mathcal{T}<\mathcal{T}_{min}$, the free energy landscape has only one global minimum at the origin and the system is stable at the thermal Minkovski spacetime enclosed by a cavity. When $\mathcal{T}_{min}<\mathcal{T}<\mathcal{T}_c$, the thermal vacuum is still globally stable although the new black hole phases emerge. Above the phase transition temperature $\mathcal{T}>\mathcal{T}_c$, the large black hole becomes globally stable. At the phase transition temperature, the thermal vacuum and the large black hole can coexist with the same free energies.

\section{Kinetics of the state switching for the asymptotically flat black holes in cavity}\label{kinetics_state_switching}

\subsection{Kinetic equation and mean first passage time}

We now discuss the effective theory of the kinetics of the state switching for the asymptotically flat black holes in a cavity in terms of the stochastic dynamics. The underlying assumption is that due to the thermal fluctuations from the reservoir, a locally stable black hole located at one basin on the free energy landscape is possible to switch its state to another local stable state located at another basin. It is also assumed that the state switching process can be coarse-grainedly described by a stochastic process of the black hole order parameter wherever the generalized free energy plays the role of the thermodynamic potential \cite{Li:2020khm,Li:2020nsy,Li:2021vdp}. 

For a quantum field theory with multiple vacuum states, the false vacuum can decay to true vacuum through either quantum tunneling \cite{Coleman:1977py,Callan:1977pt,Coleman:1980aw} or thermal fluctuations \cite{Hawking:1981fz,Weinberg:2006pc,Brown:2007sd,Oshita:2016oqn}. In this work, the cavity enclosing the black hole acts as the thermal bath. Consequently, it is expected that thermal fluctuations, rather than quantum tunneling, dominate the state switching process associated with the black hole phase transition. From the discussions in the previous two sections, we conclude that the free energy landscape provides an intuitive representation of the stability of the black hole in the cavity. Thus, the free energy landscape is a useful tool for analyzing the thermodynamics of the black hole phase transition. Furthermore, by treating the generalized free energy calculated from the Euclidean path integral approach as the effective thermodynamic potential, we can go beyond the discussion of the thermodynamics of black hole phase transition and study the kinetics of the kinetics of the state switching process.

We now outline the basic assumptions used to construct the model of the state switching process associated with the black hole phase transition under the thermal fluctuations. It is recognized that more than one branch of black hole solutions may exist, which are distinguished by their radii of the event horizon \cite{Kubiznak:2012wp,Wei:2015iwa}. Under the thermal fluctuations, the black hole may absorb or release energy in a stochastic manner, which results in an increase or decrease in the event horizon radius. Therefore, the event horizon radius is an appropriate order parameter to describe the state of the black hole, and the black hole phase transition process can be described by the stochastic process of changing the black hole radius. In phase transition theory, the order parameter is often related to a thermodynamic potential, such as the free energy or Landau-Ginzburg-Wilson free energy \cite{Landau1980}. In our case, the thermodynamic potential is the generalized free energy obtained in Eq.\eqref{Generalized_Free_energy}. Especially, it has been shown that the free energy landscape provides the gradient force for the deterministic relaxation process of fluctuating black holes \cite{Li:2022yti,Li:2024hje}. However, if only the deterministic gradient force is considered, a local stable black hole can never switch to another local or global stable black hole. Thus, we have to consider the fluctuating force or the thermal noise from the bath. In summary, we assume that the kinetics of the state switching associated with the black hole phase transition is described by the Langevin equation for the stochastic evolution of the black hole radius, where the driving forces are the gradient force from the free energy potential and the fluctuating force from the thermal bath. This kinetic model of black hole phase transition was inspired by the time-dependent Ginzburg-Landau model \cite{Hohenberg:1977ym}, where the collective behavior of the system near the critical point is described by the dissipative dynamics of a nonconserved order parameter. In general, with the free energy functional, one can set up a stochastic Langevin equation to investigate the dissipative dynamics of the system.

In analogy to the motion of a Brownian particle, the kinetics of the state switching for the asymptotically flat black hole in a cavity can be described by the Langevin equation that governs the stochastic evolution of the order parameter $x$: 
\begin{eqnarray}
    \ddot{x}+\zeta \dot{x} +\mathcal{F}'(x) -\tilde{\eta}(t)=0\;,
\end{eqnarray}
where the dot denotes the derivative with respect to the time $t$. The effective friction coefficient $\zeta$ is introduced to describe the interaction between the black hole and its reservoir. The stochastic noise $\tilde{\eta}(t)$ is Gaussian white noise. In addition, we also assume that the kinetics of the state switching for the black hole is Markovian \cite{Li:2022yti}.

In the overdamped regime, the Langevin equation can be simplified as 
\begin{eqnarray}\label{overdamped_Langevin_eq}
    \dot{x}= -\frac{1}{\zeta}\mathcal{F}'(x) +\eta(t)\;,
\end{eqnarray}
where $\eta(t)=\frac{1}{\zeta}\tilde{\eta}(t)$ is introduced for convenience. The noise with zero mean is assumed to be white Gaussianly distributed and satisfies the fluctuating-dispersion relation 
\begin{eqnarray}
    \langle \eta(t)\eta(t') \rangle=2D\delta(t-t')\;,
\end{eqnarray}
where $D=\frac{\mathcal{T}}{\zeta}$ is the diffusion coefficient. The equation \eqref{overdamped_Langevin_eq} is just the overdamped Langevin equation that describes the stochastic motion of the Brownian particle in an external potential.

We now assume that there is a large number of black hole configurations in the thermodynamic ensemble. The probability distribution of the states (on-shell solutions as well as the off-shell solutions on the free energy landscape) are denoted by $\rho(x,t)$. Then the time evolution of the probability distribution should be described by the Fokker-Planck equation which is given by 
\begin{eqnarray}\label{FPeq}
\frac{\partial \rho(x,t)}{\partial t}=D \frac{\partial}{\partial x}\left\{
e^{-\tilde \beta \mathcal{F}(x)}\frac{\partial}{\partial x}\left[e^{\tilde \beta \mathcal{F}(x)}\rho(x,t)\right]
\right\}=\mathcal{D}\rho(x,t) \;,
\end{eqnarray}
where the inverse temperature is defined as $\tilde \beta=\frac{1}{\mathcal{T}}$ and $\mathcal{D}$ is the Fokker-Planck operator. Without the loss of generality, we take $\zeta=1$ in the following.

\begin{figure}
  \centering
  \includegraphics[width=8cm]{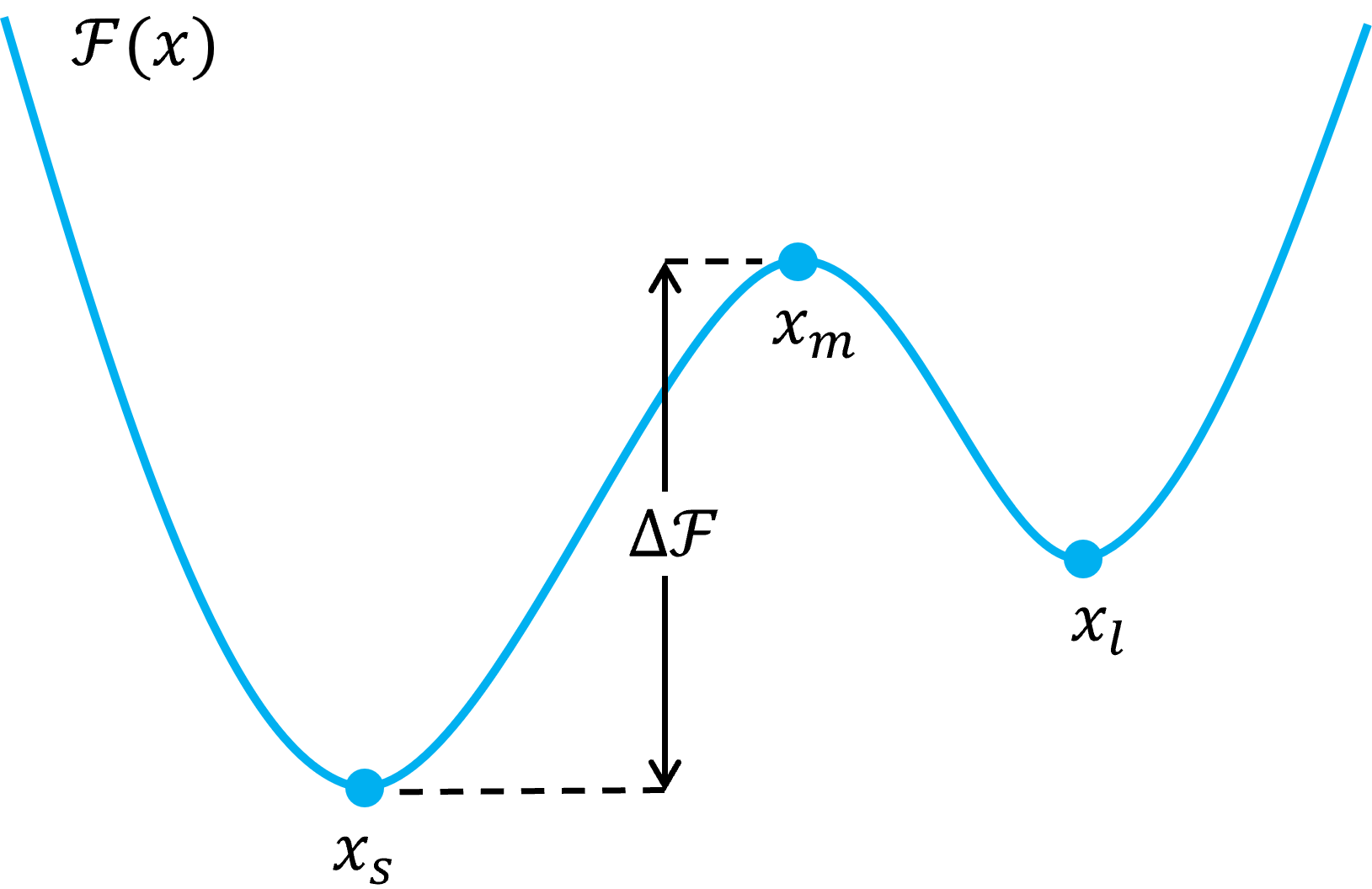}\\
  \caption{An illustration of free energy landscape with the double well shape. The order parameters of the small, the intermediate and the large black holes are denoted by $x_s$, $x_m$ and $x_l$ respectively. }
  \label{MFPT_landscape}
\end{figure}

As shown in the last section, we know that the free energy landscape as a function of the order parameter $x$ exhibits double well shape when the temperature lies in the range $[\mathcal{T}_{min}, \mathcal{T}_{max}]$. Due to the thermal fluctuations, one local stable black hole state in a potential well can make a transition to another locally stable state by going through the potential barrier (refer to Figure \ref{MFPT_landscape} for an illustration of a typical free energy landscape for such a state switching process). This state switching process can be properly characterized by the MFPT, which gives the average time scale of this stochastic event to take place for the first time \cite{Zwanzig2001}. We now study the kinetics of the black hole state switching by computing the MFPT.

As an example, we consider the state switching process from the small black hole to the large black hole. Imagine that there is a cloud of the initial black hole states located at the left well of the landscape. The black hole state will be removed from the system if it crosses the intermediate black hole state at the landscape barrier for the first time. This can be done by imposing an absorbing boundary condition at $x_m$. Then the solution to the Fokker-Planck equation in the range $x_{min}\leq x\leq x_m$ with $x_{min}=q^2+p^2$ being the lower bound of the order parameter can be formally given by 
\begin{eqnarray}
 \rho(x,t)=e^{t\mathcal{D}} \delta(x-x_0)\;,
\end{eqnarray}
where the $\delta$-function gives the initial condition with $x_{min}\leq x_0\leq x_m$. Due to the absorbing boundary condition, this probability is not conserved but decays to zero at very late time.

Define $\Sigma(x_0,t)$ to be the probability that the initial state has not made a first passage process by time $t$. Then it is given by
\begin{eqnarray}
    \Sigma(x_0,t)=\int_{x_{min}}^{x_m} \rho(x,t) dx= \int_{x_{min}}^{x_m} e^{t\mathcal{D}} \delta(x-x_0) dx\;.
\end{eqnarray}
This also vanishes at very late time. Note that the first passage time is a random variable. Its distribution function can be given by
\begin{eqnarray}
    P_F(t)=-\frac{d\Sigma(x_0,t)}{dt}\;.
\end{eqnarray}

The $n$-th moments of the FPT distribution function can be calculated from the following relation
\begin{eqnarray}\label{t_n_aver}
    \langle t^n \rangle=\int_0^{\infty} t^n P_F(t) dt=-\int_0^{\infty} t^n \frac{d\Sigma(x_0,t)}{dt} dt=n\int_0^{\infty} t^{n-1} \Sigma(x_0,t) dt \;,
\end{eqnarray}
where $n\geq 1$. By using the formal solution of Fokker-Planck equation, it can be further expressed as 
\begin{eqnarray}\label{t_n_aver_1}
    \langle t^n \rangle&=& n\int_0^{\infty} dt~ t^{n-1}   \int_{x_{min}}^{x_m} e^{t\mathcal{D}} \delta(x-x_0) dx\nonumber\\
    &=&n\int_0^{\infty} dt~ t^{n-1}   \int_{x_{min}}^{x_m} \delta(x-x_0) \left(e^{t\mathcal{D}^\dagger} 1\right)  dx\nonumber\\
    &=&n\int_0^{\infty} dt~ t^{n-1} \left(e^{t\mathcal{D}^\dagger} 1\right)\;.
\end{eqnarray}
In deriving this result, we have used the adjoint operator 
\begin{eqnarray}
\mathcal{D}^\dagger=D e^{\tilde \beta \mathcal{F}(x)} \frac{\partial}{\partial x}\left\{ e^{-\tilde \beta \mathcal{F}(x)}
\frac{\partial}{\partial x} \right\}\;, 
\end{eqnarray}
which is defined by 
\begin{eqnarray}
    \int dx \phi(x)\mathcal{D}\psi(x)=\int \psi(x)\mathcal{D}^\dagger\phi(x)\;.
\end{eqnarray}
Note that in Eq.\eqref{t_n_aver_1}, the exponential of the adjoint operates on the number $1$. Because we have performed the integration over $x$ with the delta function, the result shows that $\langle t^n \rangle$ is inherently a function of $x_0$. Then we can drop the subscript "0" and treat $\langle t^n \rangle$ as the function of $x$.

Now, applying the adjoint operator on Eq.\eqref{t_n_aver_1}, one can get the following recurrence relation 
\begin{eqnarray}
    \mathcal{D}^\dagger  \langle t^n \rangle=
    \begin{cases}
-1 & n=1 \\
-n~ \langle t^{n-1} \rangle & n\geq 2 
\end{cases}
\end{eqnarray}

When $n=1$, integrating the adjoint equation $\mathcal{D}^\dagger  \langle t \rangle=-1 $ and imposing the reflecting boundary condition at $x_{min}$, one can get the analytical expression of the MFPT for the state switching from the small black hole to the large black hole as
\begin{eqnarray}\label{MFPTexp1}
\langle t \rangle&=&\frac{1}{D}\int_{x_s}^{x_m}dx \int_{x_{min}}^{x}dx'  e^{\tilde \beta  \left(\mathcal{F}(x)-\mathcal{F}(x')\right)}\;.
\end{eqnarray}
Similarly, the analytical expression of the MFPT for the state switching from the large black hole to the small black hole is given as 
\begin{eqnarray}\label{MFPTexp2}
\langle t \rangle&=&\frac{1}{D}\int_{x_m}^{x_l}dx \int_{x}^{1}dx'  e^{\tilde \beta  \left( \mathcal{F}(x)-\mathcal{F}(x')\right)}\;.
\end{eqnarray}
Note that in deriving this expression, we have also imposed the reflecting boundary condition at the upper bound of the order parameter $x=1$.

We can also derive the analytical expression of $\langle t^2 \rangle$ from the recurrence relation. For the state switching from the small black hole to the large black hole, it is given by  
\begin{eqnarray}\label{MFPT2exp1}
\langle t^2 \rangle&=&\frac{2}{D^2}\int_{x_s}^{x_m}dx \int_{x_{min}}^{x}dx'
\int_{x'}^{x_m}dx'' \int_{x_{min}}^{x''}dx'''
e^{\tilde \beta  \left[\mathcal{F}(x)-\mathcal{F}(x')+\mathcal{F}(x'')-\mathcal{F}(x''')\right]}\;.
\end{eqnarray}
For the inverse process, it is given by 
\begin{eqnarray}\label{MFPT2exp2}
\langle t^2 \rangle&=&\frac{2}{D^2}\int_{x_m}^{x_l}dx \int_{x}^{1}dx'
\int_{x_m}^{x'}dx'' \int_{x''}^{1}dx'''
e^{\tilde \beta  \left[\mathcal{F}(x)-\mathcal{F}(x')+\mathcal{F}(x'')-\mathcal{F}(x''')\right]}\;.
\end{eqnarray}

These analytical expressions are also valid for the uncharged black hole case, in which $x_{min}$ should be replaced by $0$ and the integrating domain should be correspondingly modified to match the state switching process from the thermal vacuum to the large black hole and its inverse process.   

\subsection{Charged dyonic black holes}

We now present the numerical results of the kinetic times for the charged dyonic black holes. In the following, we set $P=0.1$ and $Q=0.15$. We are mainly interested in the dependences of the mean first passage time $\langle t \rangle$ and its relative fluctuation on the ensemble temperature. The relative fluctuation of the first passage time is defined as $\left(\langle t^2\rangle-\langle t\rangle^2 \right)/\langle t\rangle^2$, which reflects the relative deviation of the first passage time from its average value. The temperature range is taken to be $[\mathcal{T}_{min}, \mathcal{T}_{max}]$, in which the free energy landscape has the shape of double well.

\begin{figure}
  \centering
  \includegraphics[width=7cm]{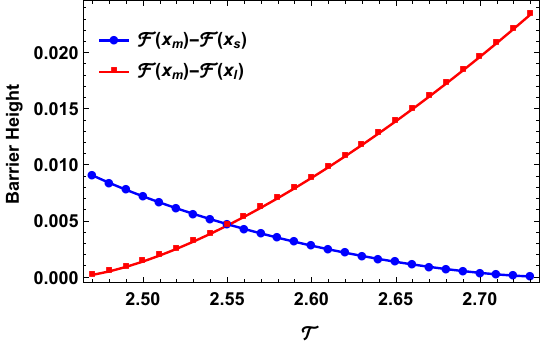}
  \caption{Barrier heights on the free energy landscapes in Figure \ref{Landscape} as the functions of the ensemble temperature.}
  \label{barrier_height}
\end{figure}

Note that the black hole state switching is essentially described by the uphill process of a particle on the free energy landscape. Therefore, the barrier height on the free energy landscape is closely related to the kinetics of the black hole state switching \cite{Li:2020khm,Li:2020nsy}. In Figure \ref{barrier_height}, we present the barrier heights $\mathcal{F}(x_m)-\mathcal{F}(x_s)$ and $\mathcal{F}(x_m)-\mathcal{F}(x_l)$ as the functions of the ensemble temperature $\mathcal{T}$. It shows that when the temperature increases, the barrier height $\mathcal{F}(x_m)-\mathcal{F}(x_s)$ is monotonic decreasing while the barrier height $\mathcal{F}(x_m)-\mathcal{F}(x_l)$ is monotonic increasing. These behaviors for the barrier heights imply that, as the temperature increases, the state switching from the small dyonic black hole to the large black hole becomes easier, whereas the inverse process becomes more difficult. In the following, we will mainly study the behaviors of kinetic times and their relation to the barrier heights.

\begin{figure}
  \centering
  \includegraphics[width=7cm]{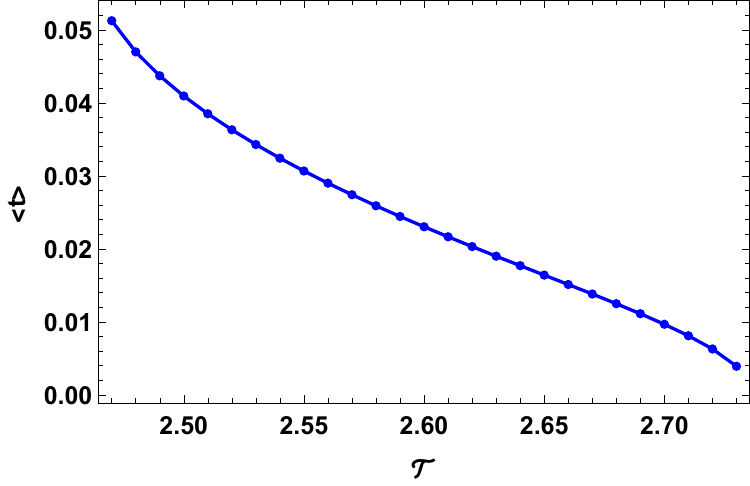}
  \includegraphics[width=7cm]{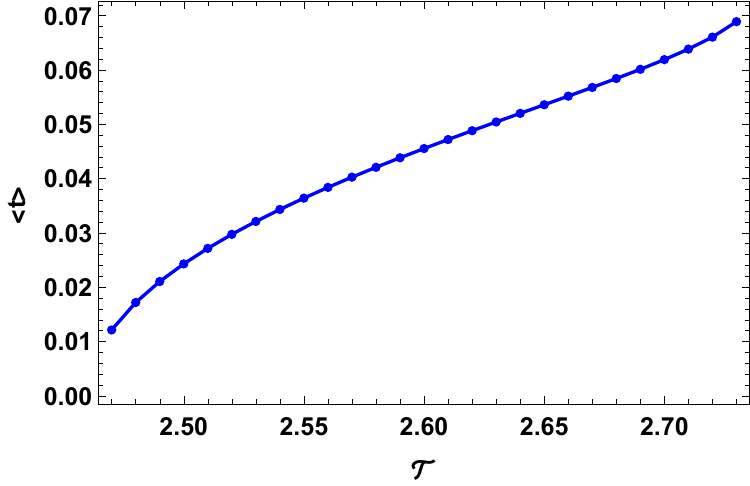}
  \caption{The dependences of mean first passage times on the ensemble temperature. The left panel is plotted for the state switching process from the small dyonic black hole to the large black hole and the right panel is for the inverse process. }
  \label{MFPT_charged}
\end{figure}

In fact, the kinetics of the black hole state switching characterized by the mean first passage time is positively correlated to the barrier height. The behaviors of the barrier heights indicate that the mean first passage time for the state switching process from the small dyonic black hole to the large black hole is the monotonic decreasing function of the ensemble temperature while the means first passage time for the inverse process is the monotonic increasing function of the ensemble temperature. This conclusion can be explicitly observed from the numerical results presented in Figure \ref{MFPT_charged}, which shows the dependences of the mean first passage times on the ensemble temperature. The plots in Figure \ref{MFPT_charged} show that, as the temperature increases,  the mean first passage time for the state switching process from the small black hole to the large black hole becomes shorter, while the kinetic time for the reverse process becomes longer. This is the expected behavior based on the analysis of the barrier heights.

\begin{figure}
  \centering
    \includegraphics[width=7cm]{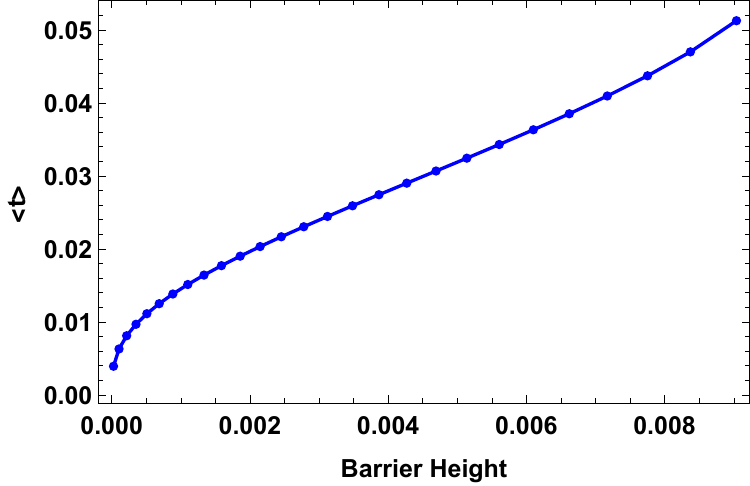}
  \includegraphics[width=7cm]{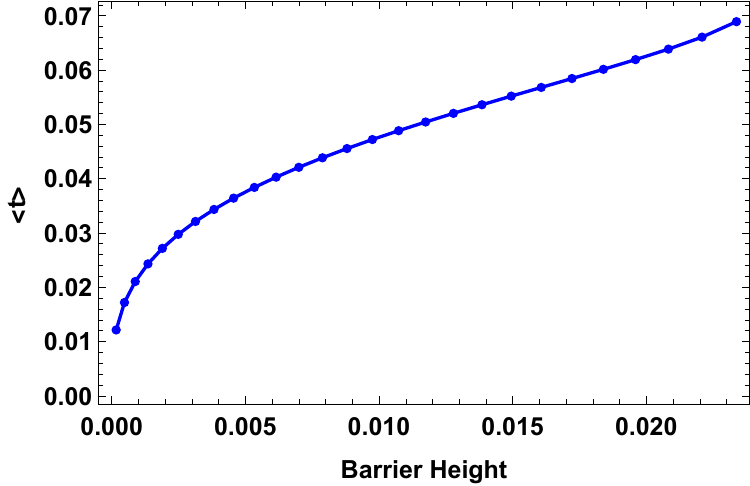}
  \caption{The correlation between the mean kinetic time and the barrier height for the state switching from the small black hole to the large black hole (left panel) and the inverse process (right panel).  }
  \label{Barrier_height_MFPT}
\end{figure}

There is another factor that can influence the kinetics of the black hole state switching process. The black hole state switching is described by the overdamped Langevin equation, which is essentially the diffusion process caused by the thermal fluctuations. These thermal fluctuations become stronger at higher temperatures. However, the behavior of mean first passage time for the process from the large black hole to the small black hole indicates that the free energy barrier is the dominant factor that impacts the kinetic time of the state switching process rather than the thermal fluctuations or the ensemble temperature. In Figure \ref{Barrier_height_MFPT}, we show the positive correlation between the mean first passage time and the barrier height explicitly. The plots show that, as long as the barrier height increases, the kinetic time correspondingly increases. It can be concluded that the free energy depths give the thermodynamic stability and the free energy landscape topography in terms of the barrier heights quantifies the mean first passage time. 

\begin{figure}
  \centering
    \includegraphics[width=7cm]{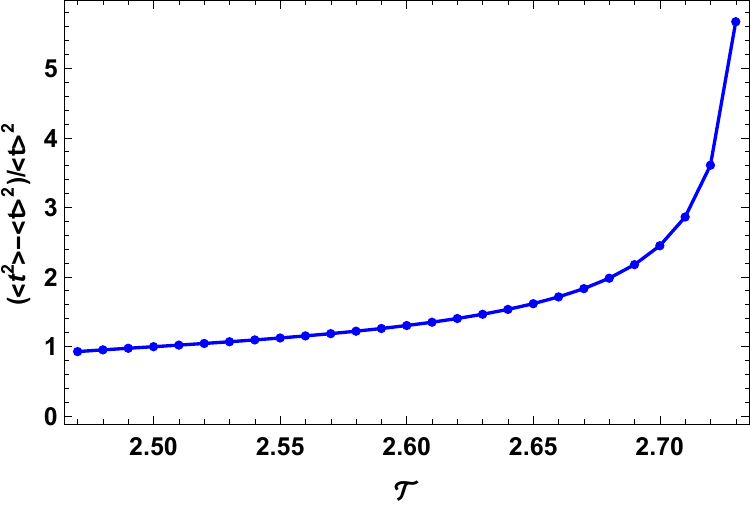}
  \includegraphics[width=7cm]{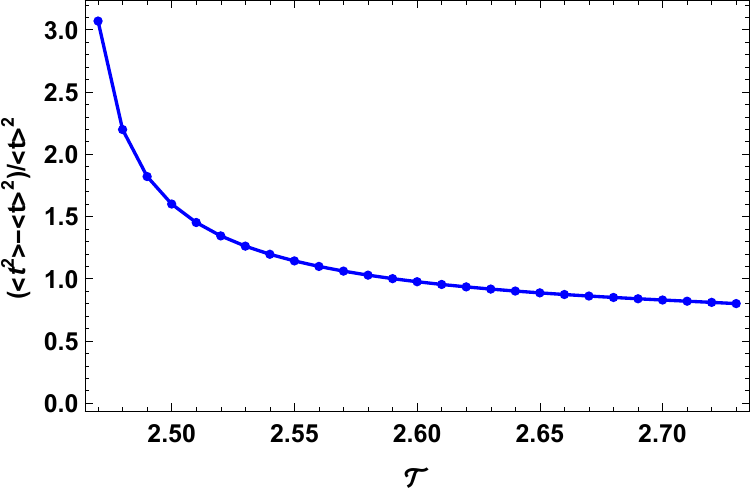}
  \caption{The dependences of the relative fluctuations on the ensemble temperature. The left panel is plotted for the state switching process from the small dyonic black hole to the large black hole and the right panel is for the inverse process. }
  \label{Rel_Fluc_charged}
\end{figure}

In Figure \ref{Rel_Fluc_charged}, we also present the numerical results of the relative fluctuations of the first passage times for the two switching processes. It shows completely opposite behaviors compared with the mean first passage times. For the small black hole to the large black hole process, the relative fluctuation becomes larger at higher temperature, while for the inverse process, it is monotonic decreasing along with the ensemble temperature. The results indicate that for the previous process, the higher ensemble temperature and the small barrier height lead to a large fluctuation, while for the latter process, the barrier height is the dominant factor.

\subsection{Uncharged black holes}

We now consider the kinetic times of state switching for the uncharged black holes. The temperature range is taken to guarantee that there are three emerged spacetime states on the free energy landscape as shown in Figure \ref{Landscape_uncharged}.

\begin{figure}
  \centering
  \includegraphics[width=7cm]{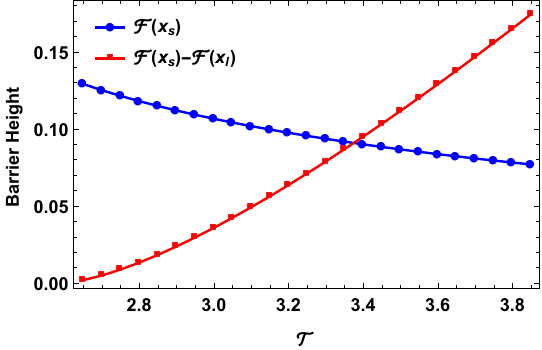}
  \caption{Barrier heights on the free energy landscapes in Figure \ref{Landscape_uncharged} as the functions of the ensemble temperature..}
  \label{Barrier_height_uncharged}
\end{figure}

In Figure \ref{Barrier_height_uncharged}, we plot the barrier heights $\mathcal{F}(x_s)$ and $\mathcal{F}(x_s)-\mathcal{F}(x_l)$ as the functions of the ensemble temperature. Note that in the present case, the small black hole is located at the potential barrier on the free energy landscape. Since the free energy of the thermal Minkovski spacetime is zero, the barrier height for the state switching process from the thermal Minkovski spacetime to the large black hole is given by the free energy of the small black hole. The dependences of the barrier heights on the ensemble temperature are shown to be similar with that for the charged dyonic black holes discussed previously.

\begin{figure}
  \centering
  \includegraphics[width=7cm]{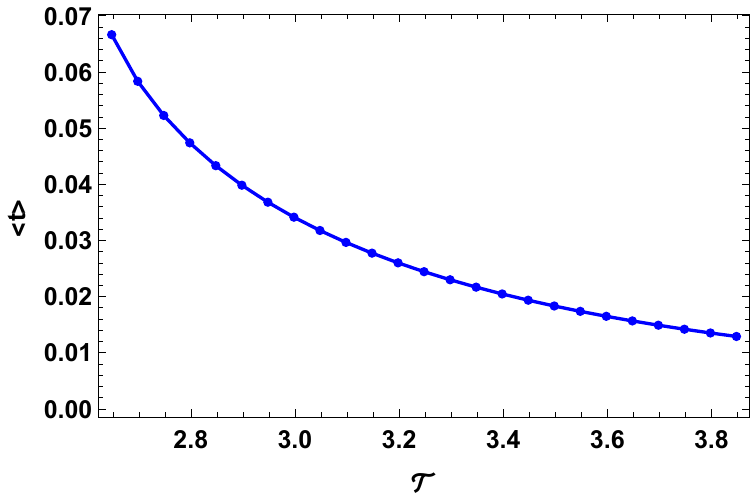}
  \includegraphics[width=7cm]{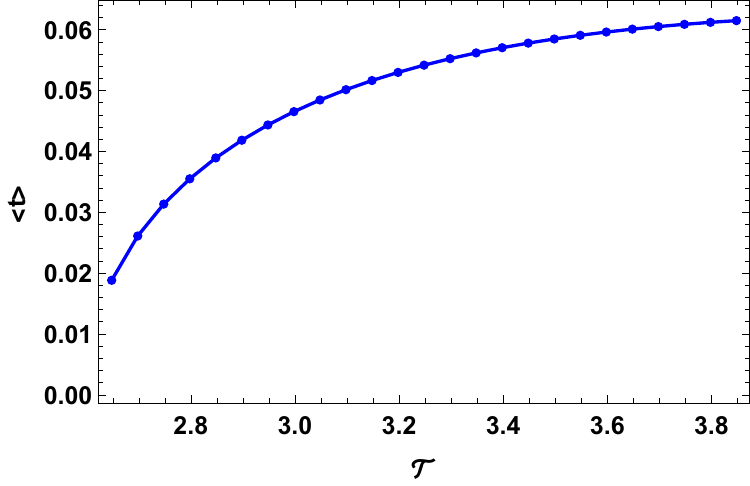}
  \caption{The dependences of mean first passage times on the ensemble temperature. The left panel is plotted for the state switching process from the small dyonic black hole to the large black hole and the right panel is for the inverse process. }
  \label{MFPT_uncharged}
\end{figure}

\begin{figure}
  \centering
  \includegraphics[width=7cm]{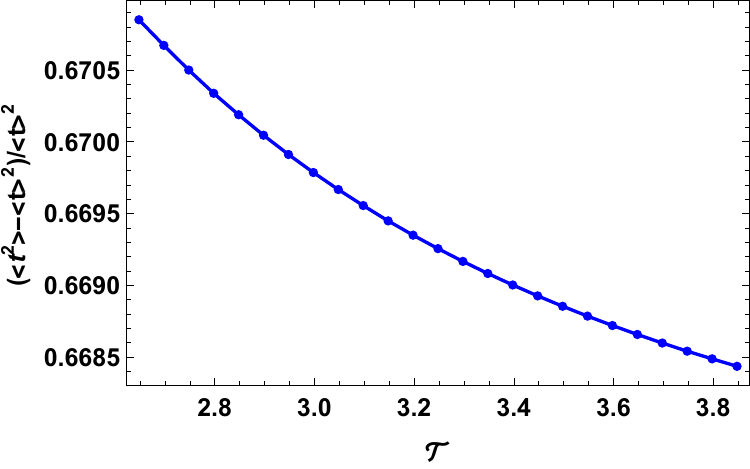}
  \includegraphics[width=7cm]{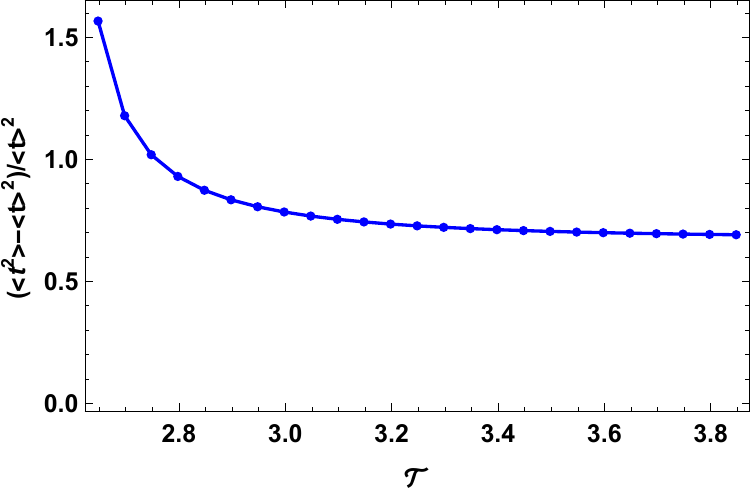}
  \caption{The dependences of the relative fluctuations on the ensemble temperature. The left panel is plotted for the state switching process from the small dyonic black hole to the large black hole and the right panel is for the inverse process.}
  \label{Rel_fluc_uncharged}
\end{figure}

In Figure \ref{MFPT_uncharged} and \ref{Rel_fluc_uncharged}, we present the numerical results of the kinetic times of the state switching processes between the thermal Minkovski spacetime and the uncharged large black hole. As discussed previously, although there are two factors that can influence the kinetics of black hole state switching, the dominant factor for the mean first passage time is the corresponding barrier height. The plots in Figure \ref{Barrier_height_uncharged} and \ref{MFPT_uncharged} show a positively correlation between the mean first passage time and the barrier height. The relative fluctuations for the uncharged black holes plotted in Figure \ref{Rel_fluc_uncharged} exhibit different behavior compared with its dependences on the temperature for the charged dyonic black holes. For the state switching process from the thermal Minkovski spacetime to the large black hole, the relative fluctuation is the monotonic decreasing function of the ensemble temperature, although its variation range is very narrow. The observation indicates that in this process the relative fluctuation is mainly dominated by the barrier height on the landscape and the influence from the ensemble temperature can be neglected. For the inverse process, the result presented in the right panel of Figure \ref{Rel_fluc_uncharged} shows the relative fluctuation is the monotonic decreasing function of the ensemble temperature and approaches a constant value at high temperature. Therefore, we can conclude that the previous mentioned two factors the ensemble temperature and the barrier height on the landscape have no effect on the relative fluctuation of the state switching process from the large black hole to the thermal Minkovski spacetime when the temperature is very high.

\section{Conclusion and discussion}
\label{con_dis}

In summary, we have employed the free energy landscape formalism to study the thermodynamics and the kinetics of state switching for the asymptotically flat black hole enclosed by a cavity. The generalized free energy for the black hole enclosed by a cavity with both the electric charge and the magnetic charge in the canonical ensemble is derived by using the York's approach, where the temperature on the cavity and the charges inside the cavity are kept as the fixed parameters. In this approach, the Euclidean manifold is regular but still admit an arbitrary ensemble temperature. It is shown that the York's approach to the black hole thermodynamics provides a natural way to study the kinetics of the state switching for the asymptotically flat black holes since the off-shell black holes are admitted to exist in the thermodynamic ensemble.

Compared with the approach used in our previous work \cite{Li:2022oup,Li:2023men} on the generalized free energies for the asymptotically AdS black holes, the introduction of an arbitrary ensemble temperature leads to a conical singularity at the event horizon of the Euclidean manifold. In this approach, the off-shell black holes refer to the Euclidean geometries with the conical singularity at the horizon. Although the specific metric used ensures that the Einstein equations are satisfied everywhere except at the conical singularity, the singularity itself represents a "delta source" of matter. One might intuitively speculate that such a geometry would not serve as a saddle point of the action functional. However, the argument presented in \cite{Susskind:1994sm} contradicts this assumption. It is noted that fixing the horizon radius when evaluating the action functional introduces a Lagrange multiplier and an associated energy density along the $r = r_+$ surface in the phase space. Consequently, the solution to the classical equations of motion derived from this action functional will inherently feature a conical singularity at the horizon. Therefore, as long as the constraint of a constant horizon radius is maintained, the Euclidean geometry with a conical singularity is indeed a stationary point of the action functional.

Within this setup, the corresponding gravitational action for the asymptotically flat charged black hole is calculated in Appendix \ref{conical_action}. This calculation demonstrates that the two approaches yield the same result when considering the redshift of the ensemble temperature in the conical singularity approach. However, there is a fundamental conceptual difference between the two methods. In our approach \cite{Li:2022oup,Li:2023men}, the off-shell black holes on the free energy landscape are described by Euclidean geometries with a conical singularity. In contrast, in York's approach \cite{York:1986it,Whiting:1988qr,Braden:1990hw}, the off-shell black holes are geometries that do not satisfy the full Einstein equations but still meet certain gravitational or electromagnetic constraints. Therefore, the two approaches produce the same form of the generalized free energy landscape with completely different interpretations.

The free energy landscape illustrates the potential pathways for gravitational phase transitions, with the off-shell black holes representing fluctuating configurations generated by thermal noise. Therefore, the difference in the physical interpretation of the off-shell black holes can be treated as representing two distinct pathways through which the state switching processes occur.

Based on the free energy landscape, we also discussed the phase transition thermodynamics of the charged dyonic black hole and the uncharged black hole respectively. It is shown that the stability of the black hole can be quantified by the topography of the free energy landscape. We also obtained the phase diagrams for the black holes in cavity, which reveal a Hawking-Page type transition for the uncharged black hole and a Van der Waals type transition for the charged black hole. Finally, we employed the stochastic Langevin equation and the corresponding Fokker-Planck equation to study the kinetics of the black hole state switching. The first passage problem for the black hole state switching was dressed analytically, where a recurrence relation for the $n$-momentum of the first passage time distribution function was obtained. This enables the analytical expressions of the kinetic times characterized by the mean first passage time and its relative fluctuation. Our numerical analysis illustrates that the kinetics of black hole state switching is determined by the ensemble temperature and the barrier height on the free energy landscape.

\appendix
\section{Euler characteristic number for 2-dimensional disk} 
\label{Euler_2d}

In this appendix, we discuss how to obtain the regularity condition Eq.\eqref{bc1_inner} for the 2-dimensional disk by calculating its Euler characteristic number.

The $\tau-r$ sector of the spacetime manifold $\mathcal{M}$ is a 2-dimensional disk $D$, which is described by the metric 
\begin{eqnarray}
    ds_D^2=b^2 d\tau^2+a^{-2}dr^2\;.
\end{eqnarray}
However, we should impose the regularity condition to guarantee that it is really a disk without singularity. The disk is bounded by $r=r_B$ where the cavity is located at. In two dimensions, the Euler characteristic number is given by 
\begin{eqnarray}
    \chi=\frac{1}{4\pi} \int_D d^2x \sqrt{g_D} R_D+\frac{1}{2\pi} \int_{\partial D} ds  k\;,
\end{eqnarray}
where $R_D$ is the scalar curvature of the disk $D$, $k$ is the geodesic curvature on the boundary $\partial D$ and $ds$ is the proper distance of the boundary. It is obvious that $ds=bd\tau$. The geodesic curvature $k$ is defined as 
\begin{eqnarray}
    k=t^\alpha n_\beta \nabla_\alpha t^\beta\;,
\end{eqnarray}
where $t^\alpha$ and $n^\alpha$ are the tangent vector and the normal vector of the boundary. For our case, they are given by 
\begin{eqnarray}
    t^\alpha=\left(\frac{1}{b},0\right)\;,\;\;\;
    n_{\alpha}=\left(0,-\frac{1}{a} \right)\;.
\end{eqnarray}
Then it is easy to obtain the geodesic curvature of the boundary as 
\begin{eqnarray}
    k=\left.\frac{ab'}{b}\right|_{r=r_B}\;.
\end{eqnarray}
By assuming that the disk is regular at the horizon $r=r_+$, one can get that the scalar curvature is given by 
\begin{eqnarray}
    R=-2\frac{a}{b}(ab')'\;.
\end{eqnarray}
Then, the Euler number is given by
\begin{eqnarray}
    \chi&=&-\int_{r_+}^{r_B} dr (ab')'+\left.(ab')\right|_{r=r_B} \nonumber\\
    &=& \left.(ab')\right|_{r=r_+}\;.
\end{eqnarray}
Due to the fact that the Euler characteristic number of a two dimensional disk is $1$, we can get the regularity condition at the horizon as
\begin{eqnarray}\label{Regular_con_app}
    \left.(ab')\right|_{r=r_+}=1\;.
\end{eqnarray}
This is just the inner boundary condition given in Eq.\eqref{bc1_inner}.
     
\section{Euler characteristic number for 4-D manifold with boundary} 
\label{Euler_4d}

In this appendix, we briefly present the result of the Euler number for the four dimensional manifold $\mathcal{M}$ with the boundary, which in turn gives the boundary condition Eq.\eqref{bc2_inner}.

For the 4-dimensional Riemann manifold with boundary, the Gauss-Bonnet-Chern formula for the Euler characteristic number is given by \cite{Taylor:2020uwf}
\begin{eqnarray}
    \chi=\int_{\mathcal{M}} \Omega +\int_{\partial\mathcal{M}} n^*\Pi\;. 
\end{eqnarray}
The bulk term is given by
\begin{eqnarray}
    \int_{\mathcal{M}} \Omega&=&\frac{1}{32\pi^2}\int_{\mathcal{M}} d^4 x \sqrt{g} \left(R_{\mu\nu\lambda\rho}R^{\mu\nu\lambda\rho}-4R_{\mu\nu}R^{\mu\nu}+R^2\right)\;,
\end{eqnarray}
This is just the so-called Gauss-Bonnet term. For the four dimensional metric given by Eq.\eqref{metric_ansatz}, the Gauss-Bonnet term is given by a compact form 
\begin{eqnarray}
    R_{\mu\nu\lambda\rho}R^{\mu\nu\lambda\rho}-4R_{\mu\nu}R^{\mu\nu}+R^2
    =\frac{8a}{r^2 b} \left(ab'(-1+a^2)\right)'\;.
\end{eqnarray}
Then, by performing the integration, the bulk term is given by
\begin{eqnarray}
    \int_{\mathcal{M}} \Omega
    &=& 2 \int_{r_+}^{r_B} dr \left(ab'(-1+a^2)\right)'\nonumber\\
    &=& \left.2(1-a^2)\right|_{r=r_+}-\left.2ab'(1-a^2)\right|_{r=r_B}\;,
\end{eqnarray}
where the regularity condition given in Eq.\eqref{Regular_con_app} was used in the last step.

The boundary term is given by 
\begin{eqnarray}
   \int_{\partial\mathcal{M}} n^*\Pi&=& \frac{1}{4\pi^2} \int_{\partial\mathcal{M}} d^3 x\sqrt{h} 
    \left(R_{ijkl}K^{ik}n^{j}n^{l}-R_{ij}K^{ij}-K R_{ij}n^{i}n^{j}\right.\nonumber\\
    &&\left. +\frac{1}{2} KR+\frac{1}{3}K^3-K\textrm{Tr}(K^2)+\frac{2}{3}\textrm{Tr}(K^3)\right)\;, 
\end{eqnarray}
where $K_{ij}$ and $n^i$ are the extrinsic curvature and the normal vector of the boundary $\partial\mathcal{M}$. It should be noted that the final result is independent of the orientation of the normal vector $n^{i}$. It can be checked that the integrand in the above equation is given by $\frac{(1-a^2)ab'}{r^2 b}$. Then the boundary term can be calculated as 
\begin{eqnarray}
     \int_{\partial\mathcal{M}} n^*\Pi=\left.2ab'(1-a^2)\right|_{r=r_B}\;, 
\end{eqnarray}
which exactly cancels the last term in the bulk integral. Therefore, the Euler characteristic number for the 4-dimensional manifold with the boundary is given by 
\begin{eqnarray}
    \chi= \left.2(1-a^2)\right|_{r=r_+}\;.
\end{eqnarray}
For our case, the 4-dimensional manifold is just the product manifold of a 2-dimensional disk and a 2-dimensional sphere. By using the formula $\chi(\mathcal{M})=\chi(D)\times \chi(S^2)=2$, one can get the condition for the metric function $a$ as 
\begin{eqnarray}
    a(r_+)=0\;.
\end{eqnarray}
This is just the inner boundary condition given in Eq.\eqref{bc2_inner}.

\section{Gravitational action for the Euclidean black hole with conical singularity} 
\label{conical_action}

In this appendix, we evaluate the gravitational action \eqref{E_H_action} for the Euclidean geometry of the charged dyonic black hole in the canonical ensemble.  

By introducing the Euclidean time $\tau=it$ and imposing the arbitrary period $\tilde{\beta}$ in $\tau$, the Euclidean geometry of the dyonic metric \eqref{dyonic_metric} has a conical singularity at the event horizon. By noting that the energy-momentum tensor of the electromagnetic field in four dimensions is traceless, the Ricci scalar vanishes for the charged Euclidean geometry far away from the conical singularity. The bulk action comes from the contribution of the conical singularity \cite{Fursaev:1995ef} and the electromagnetic field, which is given by  
\begin{eqnarray}
    I_{bulk}&=&-\frac{A}{4}\left(1-\frac{\tilde{\beta}}{\beta_H}\right)+\frac{\tilde{\beta}}{2} \left(P^2-Q^2\right)\left(\frac{1}{r_+}-\frac{1}{r_b}\right)\nonumber\\
    &=& -\frac{A}{4}+\tilde{\beta}\left(\frac{M}{2}-\frac{Q^2}{r_+}\right)-\frac{\tilde{\beta}}{2}\frac{(P^2-Q^2)}{r_b}
    \;,
\end{eqnarray}
where $\beta_H=1/T_H$ is the inverse Hawking temperature and $A=4\pi r_+^2$ is the horizon area. Here, we have used the fact that in Euclidean signature, $F^2=\frac{2\left(P^2-Q^2 \right)}{r^4}$. In the last step, we have also used the Smarr relation \eqref{Smarr_eq} to simplify the expression.

The surface terms can be evaluated as 
\begin{eqnarray}
    I_{surf}=-\tilde{\beta} r_b \left(1-\sqrt{f(r_b)}\right) +\frac{3}{2}\tilde{\beta} M-\frac{3}{2}\tilde{\beta} \frac{Q^2}{r_b}-\frac{1}{2}\tilde{\beta} \frac{P^2}{r_b}+\tilde{\beta}\frac{Q^2}{r_+}\;,
\end{eqnarray}
where a gauge transformation is introduced to preserve the regularity of the potential \cite{Gibbons:1976ue}.

The total Euclidean action is then given by 
\begin{eqnarray}\label{conical_action_result}
    I_E&=&-\frac{A}{4}+2\tilde{\beta} M -\tilde{\beta} r_b \left(1-\sqrt{f(r_b)}\right)-\tilde{\beta} \frac{\left(Q^2+P^2\right)}{r_b}\nonumber\\
    &=&\tilde{\beta}r_b \sqrt{f(r_b)}\left(1-\sqrt{f(r_b)}\right)-\pi r_+^2\;. 
\end{eqnarray}
If the reshifted inverse temperature is introduced as 
\begin{eqnarray}
    \beta=\tilde{\beta} \sqrt{f(r_b)}\;,
\end{eqnarray}
the result \eqref{conical_action_result} is shown to be equivalent to that in Eq.\eqref{York_action}. Therefore, the York's approach and the conical singularity's approach produce the same results.


\bibliographystyle{JHEP}
\bibliography{biblio.bib}

\providecommand{\href}[2]{#2}\begingroup\raggedright\begin{thebibliography}{10}

\bibitem{Hawking:1975vcx}
S.~W. Hawking, {\it {Particle Creation by Black Holes}},  {\em Commun. Math. Phys.} {\bf 43} (1975) 199--220. [Erratum: Commun.Math.Phys. 46, 206 (1976)].

\bibitem{Hawking:1982dh}
S.~W. Hawking and D.~N. Page, {\it {Thermodynamics of Black Holes in anti-De Sitter Space}},  {\em Commun. Math. Phys.} {\bf 87} (1983) 577.

\bibitem{Chamblin:1999hg}
A.~Chamblin, R.~Emparan, C.~V. Johnson, and R.~C. Myers, {\it {Holography, thermodynamics and fluctuations of charged AdS black holes}},  {\em Phys. Rev. D} {\bf 60} (1999) 104026, [\href{http://arxiv.org/abs/hep-th/9904197}{{\tt hep-th/9904197}}].

\bibitem{Chamblin:1999tk}
A.~Chamblin, R.~Emparan, C.~V. Johnson, and R.~C. Myers, {\it {Charged AdS black holes and catastrophic holography}},  {\em Phys. Rev. D} {\bf 60} (1999) 064018, [\href{http://arxiv.org/abs/hep-th/9902170}{{\tt hep-th/9902170}}].

\bibitem{Kubiznak:2012wp}
D.~Kubiznak and R.~B. Mann, {\it {P-V criticality of charged AdS black holes}},  {\em JHEP} {\bf 07} (2012) 033, [\href{http://arxiv.org/abs/1205.0559}{{\tt arXiv:1205.0559}}].

\bibitem{Altamirano:2013uqa}
N.~Altamirano, D.~Kubiz\v{n}\'ak, R.~B. Mann, and Z.~Sherkatghanad, {\it {Kerr-AdS analogue of triple point and solid/liquid/gas phase transition}},  {\em Class. Quant. Grav.} {\bf 31} (2014) 042001, [\href{http://arxiv.org/abs/1308.2672}{{\tt arXiv:1308.2672}}].

\bibitem{Wei:2019uqg}
S.-W. Wei, Y.-X. Liu, and R.~B. Mann, {\it {Repulsive Interactions and Universal Properties of Charged Anti\textendash{}de Sitter Black Hole Microstructures}},  {\em Phys. Rev. Lett.} {\bf 123} (2019), no.~7 071103, [\href{http://arxiv.org/abs/1906.10840}{{\tt arXiv:1906.10840}}].

\bibitem{Wei:2022dzw}
S.-W. Wei, Y.-X. Liu, and R.~B. Mann, {\it {Black Hole Solutions as Topological Thermodynamic Defects}},  {\em Phys. Rev. Lett.} {\bf 129} (2022), no.~19 191101, [\href{http://arxiv.org/abs/2208.01932}{{\tt arXiv:2208.01932}}].

\bibitem{Kastor:2009wy}
D.~Kastor, S.~Ray, and J.~Traschen, {\it {Enthalpy and the Mechanics of AdS Black Holes}},  {\em Class. Quant. Grav.} {\bf 26} (2009) 195011, [\href{http://arxiv.org/abs/0904.2765}{{\tt arXiv:0904.2765}}].

\bibitem{Dolan:2010ha}
B.~P. Dolan, {\it {The cosmological constant and the black hole equation of state}},  {\em Class. Quant. Grav.} {\bf 28} (2011) 125020, [\href{http://arxiv.org/abs/1008.5023}{{\tt arXiv:1008.5023}}].

\bibitem{Dolan:2011xt}
B.~P. Dolan, {\it {Pressure and volume in the first law of black hole thermodynamics}},  {\em Class. Quant. Grav.} {\bf 28} (2011) 235017, [\href{http://arxiv.org/abs/1106.6260}{{\tt arXiv:1106.6260}}].

\bibitem{York:1986it}
J.~W. York, Jr., {\it {Black hole thermodynamics and the Euclidean Einstein action}},  {\em Phys. Rev. D} {\bf 33} (1986) 2092--2099.

\bibitem{Whiting:1988qr}
B.~F. Whiting and J.~W. York, Jr., {\it {Action Principle and Partition Function for the Gravitational Field in Black Hole Topologies}},  {\em Phys. Rev. Lett.} {\bf 61} (1988) 1336.

\bibitem{Braden:1990hw}
H.~W. Braden, J.~D. Brown, B.~F. Whiting, and J.~W. York, Jr., {\it {Charged black hole in a grand canonical ensemble}},  {\em Phys. Rev. D} {\bf 42} (1990) 3376--3385.

\bibitem{Carlip:2003ne}
S.~Carlip and S.~Vaidya, {\it {Phase transitions and critical behavior for charged black holes}},  {\em Class. Quant. Grav.} {\bf 20} (2003) 3827--3838, [\href{http://arxiv.org/abs/gr-qc/0306054}{{\tt gr-qc/0306054}}].

\bibitem{Lundgren:2006kt}
A.~P. Lundgren, {\it {Charged black hole in a canonical ensemble}},  {\em Phys. Rev. D} {\bf 77} (2008) 044014, [\href{http://arxiv.org/abs/gr-qc/0612119}{{\tt gr-qc/0612119}}].

\bibitem{Basu:2016srp}
P.~Basu, C.~Krishnan, and P.~N. Bala~Subramanian, {\it {Hairy Black Holes in a Box}},  {\em JHEP} {\bf 11} (2016) 041, [\href{http://arxiv.org/abs/1609.01208}{{\tt arXiv:1609.01208}}].

\bibitem{Brown:1994gs}
J.~D. Brown, J.~Creighton, and R.~B. Mann, {\it {Temperature, energy and heat capacity of asymptotically anti-de Sitter black holes}},  {\em Phys. Rev. D} {\bf 50} (1994) 6394--6403, [\href{http://arxiv.org/abs/gr-qc/9405007}{{\tt gr-qc/9405007}}].

\bibitem{Peca:1998cs}
C.~S. Peca and J.~P.~S. Lemos, {\it {Thermodynamics of Reissner-Nordstrom anti-de Sitter black holes in the grand canonical ensemble}},  {\em Phys. Rev. D} {\bf 59} (1999) 124007, [\href{http://arxiv.org/abs/gr-qc/9805004}{{\tt gr-qc/9805004}}].

\bibitem{Mitra:1999ge}
P.~Mitra, {\it {Thermodynamics of charged anti-de Sitter black holes in canonical ensemble}},  {\em Phys. Lett. B} {\bf 459} (1999) 119--124, [\href{http://arxiv.org/abs/gr-qc/9903078}{{\tt gr-qc/9903078}}].

\bibitem{Wang:2001gt}
B.~B. Wang and C.~G. Huang, {\it {Thermodynamics of de Sitter space-time in York's formalism}},  {\em Mod. Phys. Lett. A} {\bf 16} (2001) 1487--1492.

\bibitem{Wang:2002nq}
B.~B. Wang and C.~G. Huang, {\it {Thermodynamics of Reissner-Nordstrom-De Sitter black hole in York's formalism}},  {\em Class. Quant. Grav.} {\bf 19} (2002) 2491--2502.

\bibitem{Simovic:2018tdy}
F.~Simovic and R.~B. Mann, {\it {Critical Phenomena of Charged de Sitter Black Holes in Cavities}},  {\em Class. Quant. Grav.} {\bf 36} (2019), no.~1 014002, [\href{http://arxiv.org/abs/1807.11875}{{\tt arXiv:1807.11875}}].

\bibitem{Wang:2019cax}
P.~Wang, H.~Wu, and H.~Yang, {\it {Thermodynamic Geometry of AdS Black Holes and Black Holes in a Cavity}},  {\em Eur. Phys. J. C} {\bf 80} (2020), no.~3 216, [\href{http://arxiv.org/abs/1910.07874}{{\tt arXiv:1910.07874}}].

\bibitem{Wang:2020hjw}
P.~Wang, H.~Wu, H.~Yang, and F.~Yao, {\it {Extended Phase Space Thermodynamics for Black Holes in a Cavity}},  {\em JHEP} {\bf 09} (2020) 154, [\href{http://arxiv.org/abs/2006.14349}{{\tt arXiv:2006.14349}}].

\bibitem{Haroon:2020vpr}
S.~Haroon, R.~A. Hennigar, R.~B. Mann, and F.~Simovic, {\it {Thermodynamics of Gauss-Bonnet-de Sitter Black Holes}},  {\em Phys. Rev. D} {\bf 101} (2020) 084051, [\href{http://arxiv.org/abs/2002.01567}{{\tt arXiv:2002.01567}}].

\bibitem{Banihashemi:2022jys}
B.~Banihashemi and T.~Jacobson, {\it {Thermodynamic ensembles with cosmological horizons}},  {\em JHEP} {\bf 07} (2022) 042, [\href{http://arxiv.org/abs/2204.05324}{{\tt arXiv:2204.05324}}].

\bibitem{Jacobson:2022gmo}
T.~Jacobson and M.~R. Visser, {\it {Entropy of causal diamond ensembles}},  {\em SciPost Phys.} {\bf 15} (2023), no.~1 023, [\href{http://arxiv.org/abs/2212.10608}{{\tt arXiv:2212.10608}}].

\bibitem{Jacobson:2022jir}
T.~Jacobson and M.~R. Visser, {\it {Partition Function for a Volume of Space}},  {\em Phys. Rev. Lett.} {\bf 130} (2023), no.~22 221501, [\href{http://arxiv.org/abs/2212.10607}{{\tt arXiv:2212.10607}}].

\bibitem{Draper:2022ofa}
P.~Draper and S.~Farkas, {\it {Euclidean de Sitter black holes and microcanonical equilibrium}},  {\em Phys. Rev. D} {\bf 105} (2022), no.~12 126021, [\href{http://arxiv.org/abs/2203.01871}{{\tt arXiv:2203.01871}}].

\bibitem{Lemos:2024sjs}
J.~P.~S. Lemos and O.~B. Zaslavskii, {\it {Hot spaces with positive cosmological constant in the canonical ensemble: de Sitter solution, Schwarzschild\textendash{}de Sitter black hole, and Nariai universe}},  {\em Phys. Rev. D} {\bf 109} (2024), no.~8 084016, [\href{http://arxiv.org/abs/2402.05166}{{\tt arXiv:2402.05166}}].

\bibitem{Lu:2010xt}
J.~X. Lu, S.~Roy, and Z.~Xiao, {\it {Phase transitions and critical behavior of black branes in canonical ensemble}},  {\em JHEP} {\bf 01} (2011) 133, [\href{http://arxiv.org/abs/1010.2068}{{\tt arXiv:1010.2068}}].

\bibitem{Lu:2013nt}
J.~X. Lu and R.~Wei, {\it {Modulating the phase structure of black D6 branes in canonical ensemble}},  {\em JHEP} {\bf 04} (2013) 100, [\href{http://arxiv.org/abs/1301.1780}{{\tt arXiv:1301.1780}}].

\bibitem{Wang:2019urm}
P.~Wang, H.~Yang, and S.~Ying, {\it {Thermodynamics and phase transition of a Gauss-Bonnet black hole in a cavity}},  {\em Phys. Rev. D} {\bf 101} (2020), no.~6 064045, [\href{http://arxiv.org/abs/1909.01275}{{\tt arXiv:1909.01275}}].

\bibitem{Peng:2020zmu}
Y.~Peng, {\it {Analytical investigations on formations of hairy neutral reflecting shells in the scalar-Gauss\textendash{}Bonnet gravity}},  {\em Eur. Phys. J. C} {\bf 80} (2020), no.~3 202, [\href{http://arxiv.org/abs/2002.01892}{{\tt arXiv:2002.01892}}].

\bibitem{Su:2021jto}
B.-Y. Su and N.~Li, {\it {On the dual relation in the Hawking\textendash{}Page phase transition of the black holes in a cavity}},  {\em Nucl. Phys. B} {\bf 979} (2022) 115782, [\href{http://arxiv.org/abs/2105.06670}{{\tt arXiv:2105.06670}}].

\bibitem{Marks:2021fpe}
G.~A. Marks, F.~Simovic, and R.~B. Mann, {\it {Phase transitions in 4D Gauss\textendash{}Bonnet\textendash{}de Sitter black holes}},  {\em Phys. Rev. D} {\bf 104} (2021), no.~10 104056, [\href{http://arxiv.org/abs/2107.11352}{{\tt arXiv:2107.11352}}].

\bibitem{Peng:2017gss}
Y.~Peng, B.~Wang, and Y.~Liu, {\it {On the thermodynamics of the black hole and hairy black hole transitions in the asymptotically flat spacetime with a box}},  {\em Eur. Phys. J. C} {\bf 78} (2018), no.~3 176, [\href{http://arxiv.org/abs/1708.01411}{{\tt arXiv:1708.01411}}].

\bibitem{Simovic:2019zgb}
F.~Simovic and R.~B. Mann, {\it {Critical Phenomena of Born-Infeld-de Sitter Black Holes in Cavities}},  {\em JHEP} {\bf 05} (2019) 136, [\href{http://arxiv.org/abs/1904.04871}{{\tt arXiv:1904.04871}}].

\bibitem{Liang:2019dni}
K.~Liang, P.~Wang, H.~Wu, and M.~Yang, {\it {Phase structures and transitions of Born\textendash{}Infeld black holes in a grand canonical ensemble}},  {\em Eur. Phys. J. C} {\bf 80} (2020), no.~3 187, [\href{http://arxiv.org/abs/1907.00799}{{\tt arXiv:1907.00799}}].

\bibitem{Lemos:1996bq}
J.~P.~S. Lemos, {\it {Thermodynamics of the two-dimensional black hole in the Teitelboim-Jackiw theory}},  {\em Phys. Rev. D} {\bf 54} (1996) 6206--6212, [\href{http://arxiv.org/abs/gr-qc/9608016}{{\tt gr-qc/9608016}}].

\bibitem{Andre:2020czm}
R.~Andr\'e and J.~P.~S. Lemos, {\it {Thermodynamics of five-dimensional Schwarzschild black holes in the canonical ensemble}},  {\em Phys. Rev. D} {\bf 102} (2020), no.~2 024006, [\href{http://arxiv.org/abs/2006.10050}{{\tt arXiv:2006.10050}}].

\bibitem{Andre:2021ctu}
R.~Andr\'e and J.~P.~S. Lemos, {\it {Thermodynamics of $d$-dimensional Schwarzschild black holes in the canonical ensemble}},  {\em Phys. Rev. D} {\bf 103} (2021), no.~6 064069, [\href{http://arxiv.org/abs/2101.11010}{{\tt arXiv:2101.11010}}].

\bibitem{Fernandes:2023byx}
T.~V. Fernandes and J.~P.~S. Lemos, {\it {Grand canonical ensemble of a d-dimensional Reissner-Nordstr\"om black hole in a cavity}},  {\em Phys. Rev. D} {\bf 108} (2023), no.~8 084053, [\href{http://arxiv.org/abs/2309.12388}{{\tt arXiv:2309.12388}}].

\bibitem{Li:2020khm}
R.~Li and J.~Wang, {\it {Thermodynamics and kinetics of Hawking-Page phase transition}},  {\em Phys. Rev. D} {\bf 102} (2020), no.~2 024085.

\bibitem{Li:2020nsy}
R.~Li, K.~Zhang, and J.~Wang, {\it {Thermal dynamic phase transition of Reissner-Nordstr\"om Anti-de Sitter black holes on free energy landscape}},  {\em JHEP} {\bf 10} (2020) 090, [\href{http://arxiv.org/abs/2008.00495}{{\tt arXiv:2008.00495}}].

\bibitem{Li:2021vdp}
R.~Li, K.~Zhang, and J.~Wang, {\it {Probing black hole microstructure with the kinetic turnover of phase transition}},  {\em Phys. Rev. D} {\bf 104} (2021), no.~8 084076, [\href{http://arxiv.org/abs/2102.09439}{{\tt arXiv:2102.09439}}].

\bibitem{Carter}
B.~Carter, {\it {Black Hole Equilibrium States, in: Black Holes (eds. C. DeWitt and B. S. DeWitt, Gordon and Breach Science Publishers, New York, 1973) p. 56.}}, .

\bibitem{Carter:2009nex}
B.~Carter, {\it {Republication of: Black hole equilibrium states}},  {\em Gen. Rel. Grav.} {\bf 41} (2009), no.~12 2873--2938.

\bibitem{Rasheed:1995zv}
D.~Rasheed, {\it {The Rotating dyonic black holes of Kaluza-Klein theory}},  {\em Nucl. Phys. B} {\bf 454} (1995) 379--401, [\href{http://arxiv.org/abs/hep-th/9505038}{{\tt hep-th/9505038}}].

\bibitem{Campbell:1992hc}
B.~A. Campbell, N.~Kaloper, R.~Madden, and K.~A. Olive, {\it {Physical properties of four-dimensional superstring gravity black hole solutions}},  {\em Nucl. Phys. B} {\bf 399} (1993) 137--168, [\href{http://arxiv.org/abs/hep-th/9301129}{{\tt hep-th/9301129}}].

\bibitem{Cheng:1993wp}
G.-J. Cheng, R.-R. Hsu, and W.-F. Lin, {\it {Dyonic black holes in string theory}},  {\em J. Math. Phys.} {\bf 35} (1994) 4839--4847, [\href{http://arxiv.org/abs/hep-th/9302065}{{\tt hep-th/9302065}}].

\bibitem{Lu:2013ura}
H.~L\"u, Y.~Pang, and C.~N. Pope, {\it {AdS Dyonic Black Hole and its Thermodynamics}},  {\em JHEP} {\bf 11} (2013) 033, [\href{http://arxiv.org/abs/1307.6243}{{\tt arXiv:1307.6243}}].

\bibitem{Dutta:2013dca}
S.~Dutta, A.~Jain, and R.~Soni, {\it {Dyonic Black Hole and Holography}},  {\em JHEP} {\bf 12} (2013) 060, [\href{http://arxiv.org/abs/1310.1748}{{\tt arXiv:1310.1748}}].

\bibitem{Jiang:2023gas}
J.~Jiang and J.~Tan, {\it {Spontaneous scalarization of dyonic black hole in Einstein-Maxwell-scalar theory}},  {\em Eur. Phys. J. C} {\bf 83} (2023), no.~4 290.

\bibitem{Gibbons:1976ue}
G.~W. Gibbons and S.~W. Hawking, {\it {Action Integrals and Partition Functions in Quantum Gravity}},  {\em Phys. Rev. D} {\bf 15} (1977) 2752--2756.

\bibitem{Coleman:1977py}
S.~R. Coleman, {\it {The Fate of the False Vacuum. 1. Semiclassical Theory}},  {\em Phys. Rev. D} {\bf 15} (1977) 2929--2936. [Erratum: Phys.Rev.D 16, 1248 (1977)].

\bibitem{Callan:1977pt}
C.~G. Callan, Jr. and S.~R. Coleman, {\it {The Fate of the False Vacuum. 2. First Quantum Corrections}},  {\em Phys. Rev. D} {\bf 16} (1977) 1762--1768.

\bibitem{Coleman:1980aw}
S.~R. Coleman and F.~De~Luccia, {\it {Gravitational Effects on and of Vacuum Decay}},  {\em Phys. Rev. D} {\bf 21} (1980) 3305.

\bibitem{Hawking:1981fz}
S.~W. Hawking and I.~G. Moss, {\it {Supercooled Phase Transitions in the Very Early Universe}},  {\em Phys. Lett. B} {\bf 110} (1982) 35--38.

\bibitem{Weinberg:2006pc}
E.~J. Weinberg, {\it {Hawking-Moss bounces and vacuum decay rates}},  {\em Phys. Rev. Lett.} {\bf 98} (2007) 251303, [\href{http://arxiv.org/abs/hep-th/0612146}{{\tt hep-th/0612146}}].

\bibitem{Brown:2007sd}
A.~R. Brown and E.~J. Weinberg, {\it {Thermal derivation of the Coleman-De Luccia tunneling prescription}},  {\em Phys. Rev. D} {\bf 76} (2007) 064003, [\href{http://arxiv.org/abs/0706.1573}{{\tt arXiv:0706.1573}}].

\bibitem{Oshita:2016oqn}
N.~Oshita and J.~Yokoyama, {\it {Entropic interpretation of the Hawking\textendash{}Moss bounce}},  {\em PTEP} {\bf 2016} (2016), no.~5 051E02, [\href{http://arxiv.org/abs/1603.06671}{{\tt arXiv:1603.06671}}].

\bibitem{Wei:2015iwa}
S.-W. Wei and Y.-X. Liu, {\it {Insight into the Microscopic Structure of an AdS Black Hole from a Thermodynamical Phase Transition}},  {\em Phys. Rev. Lett.} {\bf 115} (2015), no.~11 111302, [\href{http://arxiv.org/abs/1502.00386}{{\tt arXiv:1502.00386}}]. [Erratum: Phys.Rev.Lett. 116, 169903 (2016)].

\bibitem{Landau1980}
E.~L. L.~D.~Landau, {\em Lifshitz, Statistical Physics. Part 1}.
\newblock Pergamon Press, Oxford, 1980.

\bibitem{Li:2022yti}
R.~Li and J.~Wang, {\it {Non-Markovian dynamics of black hole phase transition}},  {\em Phys. Rev. D} {\bf 106} (2022), no.~10 104039, [\href{http://arxiv.org/abs/2205.00594}{{\tt arXiv:2205.00594}}].

\bibitem{Li:2024hje}
R.~Li, C.~Liu, and J.~Wang, {\it {Phase space path integral approach to the kinetics of black hole phase transition}},  {\em Phys. Rev. D} {\bf 110} (2024), no.~2 024079, [\href{http://arxiv.org/abs/2401.02260}{{\tt arXiv:2401.02260}}].

\bibitem{Hohenberg:1977ym}
P.~C. Hohenberg and B.~I. Halperin, {\it {Theory of Dynamic Critical Phenomena}},  {\em Rev. Mod. Phys.} {\bf 49} (1977) 435--479.

\bibitem{Zwanzig2001}
R.~Zwanzig, {\em nonequilibrium Statistical Mechanics}.
\newblock Oxford University Press, New York, 2001.

\bibitem{Li:2022oup}
R.~Li and J.~wang, {\it {Generalized free energy landscape of a black hole phase transition}},  {\em Phys. Rev. D} {\bf 106} (2022), no.~10 106015, [\href{http://arxiv.org/abs/2206.02623}{{\tt arXiv:2206.02623}}].

\bibitem{Li:2023men}
R.~Li and J.~Wang, {\it {Generalized free energy landscapes of charged Gauss-Bonnet-AdS black holes in diverse dimensions}},  {\em Phys. Rev. D} {\bf 108} (2023), no.~4 044057, [\href{http://arxiv.org/abs/2304.03425}{{\tt arXiv:2304.03425}}].

\bibitem{Susskind:1994sm}
L.~Susskind and J.~Uglum, {\it {Black hole entropy in canonical quantum gravity and superstring theory}},  {\em Phys. Rev. D} {\bf 50} (1994) 2700--2711, [\href{http://arxiv.org/abs/hep-th/9401070}{{\tt hep-th/9401070}}].

\bibitem{Taylor:2020uwf}
M.~Taylor and L.~Too, {\it {Renormalized entanglement entropy and curvature invariants}},  {\em JHEP} {\bf 12} (2020) 050, [\href{http://arxiv.org/abs/2004.09568}{{\tt arXiv:2004.09568}}].

\bibitem{Fursaev:1995ef}
D.~V. Fursaev and S.~N. Solodukhin, {\it {On the description of the Riemannian geometry in the presence of conical defects}},  {\em Phys. Rev. D} {\bf 52} (1995) 2133--2143, [\href{http://arxiv.org/abs/hep-th/9501127}{{\tt hep-th/9501127}}].

\end{thebibliography}\endgroup

\end{document}